\documentclass[12pt]{article}
\usepackage{graphicx,amssymb,amsmath,empheq}
\usepackage{authblk}
\usepackage{amsthm,bm}
\usepackage{empheq}
\usepackage{subfig} 

\usepackage{hyperref}
\hypersetup{colorlinks = true, linkcolor=black, citecolor=black, urlcolor=blue}
\usepackage{url}

\theoremstyle{plain}

\usepackage{tikz}
\usepackage{tikz-cd}
\usetikzlibrary{arrows}
\usetikzlibrary{intersections}
\usetikzlibrary{shapes.geometric}
\usetikzlibrary{decorations.pathmorphing, patterns,shapes}
\usetikzlibrary{decorations.markings}

\tikzset{
  mid arrow/.style={postaction={decorate,decoration={
        markings,
        mark=at position .575 with {\arrow[#1]{stealth}}
      }}},
  near arrow/.style={postaction={decorate,decoration={
        markings,
        mark=at position .275 with {\arrow[#1]{stealth}}
      }}},
   far arrow/.style={postaction={decorate,decoration={
        markings,
        mark=at position .800 with {\arrow[#1]{stealth}}
      }}},
}

\tikzset{
  baseline = -0.5ex,
  wavy/.style = {
    thick,
    decorate,
    decoration={snake,amplitude=2pt,segment length=5pt}},
  sdot/.style = {
    circle,
    draw=none,
    fill=black,
    minimum size=2.5pt,
    inner sep=0pt},
  bdot/.style = {
    circle,
    draw=none,
    fill=black,
    minimum size=4pt,
    inner sep=0pt},
  svertex/.style = {
    circle,
    draw=black,
    thick,
    fill=lightgray,
    minimum size=8pt,
    inner sep=1pt},
  mvertex/.style = {
    circle,
    draw=black,
    thick,
    fill=lightgray,
    minimum size=12pt,
    inner sep=1pt},
  bvertex/.style = {
    circle,
    draw=black,
    thick,
    fill=lightgray,
    minimum size=16pt}}

\usepackage[nosort]{cite}

\renewcommand{\bar}{\overline}
\renewcommand{\tilde}{\widetilde}

\renewcommand{\leq}{\leqslant}
\renewcommand{\geq}{\geqslant}

\newcommand{\dkap}{\delta\kern-1.25pt\varkappa}

\makeatletter

\newcommand*{\wideboxed}[1]{\setlength{\fboxsep}{1ex}%
  \fbox{\m@th$\displaystyle#1$}}
\makeatother

\title{
Information Scrambling and Entanglement Dynamics of Complex Brownian Sachdev-Ye-Kitaev Models
}

\author[1]{Pengfei Zhang}
\affil[1]{\normalsize \it Department of Physics, Fudan University, Shanghai, 200438, China}

\date{\today}

\begin{document}
  \maketitle
  
  \begin{abstract}
In this work, we study the information scrambling and the entanglement dynamics in the complex Brownian Sachdev-Ye-Kitaev (cBSYK) models, focusing on their dependence on the charge density $n$. We first derive the effective theory for scramblons in a single cBSYK model, which gives closed-form expressions for the late-time OTOC and operator size. In particular, the result for OTOC is consistent with numerical observations in \cite{agarwal2022emergent}. We then study the entanglement dynamics in cBSYK chains. We derive the density dependence of the entanglement velocity for both R\'enyi entropies and the Von Neumann entropy, with a comparison to the butterfly velocity. We further consider adding repeated measurements and derive the effective theory of the measurement induced transition which shows $U(2)_L\otimes U(2)_R$ symmetry for non-interacting models.
  \end{abstract}
  \tableofcontents
\section{Introduction}
Understanding the dynamics of quantum information is of vital importance for revealing the universal picture of quantum many-body systems in both high-energy physics and condensed matter physics. For example, motivated by gravity calculations, the out-of-time-order correlator (OTOC) $\langle W^\dagger(t)V^\dagger(0) W(t)V(0)\rangle $ was introduced to describe the quantum information scrambling in generali quantum systems \cite{Hayden:2007cs,Sekino:2008he,shenker2014black,kitaev2014hidden}. At the early-time regime with $t\ll t_s$, it shows an exponential deviation behavior $1-\# e^{\varkappa t}/N$ in systems with large Hilbert space dimensions, which defines the quantum Lyapunov exponent $\varkappa$. It has been proved that the quantum Lyapunov has an upper bound $\varkappa \leq 2\pi/\beta$ \cite{maldacena2016bound}, which is saturated by holographic systems dual to semi-classical black holes. The study of information scrambling largely benefits from the Sachdev-Ye-Kitaev (SYK) models \cite{kitaevsimple,sachdev1993gapless,maldacena2016remarks,kitaev2018soft}, which describes randomly interacting Majorana fermions. The SYK model can be solved using the $1/N$ expansion. The early-time OTOC in the SYK model is known to show maximally chaotic behavior in the low-temperature limit \cite{kitaevsimple,maldacena2016remarks,kitaev2018soft}, while its late-time bahavior $t\gtrsim t_s$ is described by an emergent bulk scattering with a complex phase shift \cite{gu2022two,Stanford:2021bhl}. Moreover, a series of works study the entanglement of the SYK model, where the R\'enyi entropies are computed exactly to the leading order of $1/N$ using the perturbation theory or numerical simulations \cite{Gu:2017njx,Liu:2017kfa,zhang2020entanglement,chen2020replica,Zhang:2020kia,Su:2020quk,Haldar:2020ymg,Chen:2020atj}, which is closely related to replica wormholes in gravity \cite{Penington:2019kki,Almheiri:2020cfm,almheiri2020replica}. In special limits, results for the Von Neumann can also be obtained by performing the analytical continuation \cite{Chen:2019qqe,dadras2021perturbative,jian2021phase,Dadras:2019tcz}. 

Time scales in quantum information dynamics play an important role in understanding ``Planckian'' transports in strongly correlated materials \cite{hartnoll2018holographic,hod2007universal,hartnoll2015theory,legros2019universal,lucas2019operator,blake2016universal,gu2017local}. Although the original SYK model is defined for Majorana fermions with all-to-all interactions, a number of generalizations have been proposed to study quantum transports, including models with Brownian couplings \cite{Saad:2018bqo,Sunderhauf:2019djv}, charge conservations \cite{Davison:2016ngz,Bulycheva:2017uqj,Chaturvedi:2018uov,Gu:2019jub,chen2020many}, or at higher dimensions \cite{Gu:2016oyy,Gu:2017ohj,Song:2017pfw,Chen:2017dbb,zhang2017dispersive,jian2017solvable}. As an example, authors propose bounds for the Lyapunov expoenent in systems with charge conservations motivated by the calculation in complex SYK models \cite{chen2020many}. In particular, the Lyapunov exponent and butterfly velocity are computed exactly in the complex Brownian SYK (cBSYK) model. Later, there are attempts to understand the information scrambling in the cBSYK model beyond the early-time regime \cite{agarwal2022emergent}. By generalizing the mapping from the Majorana Brownian SYK model to the effective bosonic model in \cite{Sunderhauf:2019djv}, authors are above to perform numerical simulations for the OTOC in systems with finite but large $N$, where a data collapse for OTOC with different charge density $n$ has been observed. 

Motivated by these developments, here we push forward the understanding of the quantum information dynamics in the cBSYK model by studying its information scrambling and entanglement dynamics. In section 2, we derive closed-form expressions for the late-time OTOC, which is consistent with numerical observations. We also introduce the idea of finite density operator size and compute its distribution function in the late-time regime. In section 3, we study the entanglement dynamics of cBSYK chains, including the density dependence of the entanglement velocity, and the effective theory for the measurement induced transitions
. We conclude our paper in section 4, with discussions for a few future directions.

\section{Quantum Information Scrambling at Late Time}
In this section, we study the information scrambling of the cBSYK model, focusing on its dependence on charge density $n$. Adapting the methodology developed in \cite{gu2022two,sizenewpaper,Zhang:2022knu}, we derive analytical results for the late-time OTOC and operator size distribution by summing up contributions with multiple scramblons. In particular, our results explain the $n$ dependence of OTOC observed in recent numerics \cite{agarwal2022emergent}.

\subsection{Model and two-point functions}\label{cBSYKsub1}
The complex Brownian SYK model with $q$-body interactions (cBSYK$_q$) is described by the Hamiltonian:
\begin{equation}\label{eqn:cBSYKH}
H(t)=\sum_{i_1<i_2...<i_{\frac{q}{2}}}\sum_{j_1<j_2...<j_{\frac{q}{2}}}J_{i_1i_2...i_{\frac{q}{2}}j_1j_2...j_{\frac{q}{2}}}(t)c^\dagger_{i_1}c^\dagger_{i_2}...c^\dagger_{i_{\frac{q}{2}}}c^{}_{j_1}c^{}_{j_2}...c^{}_{j_{\frac{q}{2}}}.
\end{equation}
Here the random interaction parameters $J_{i_1i_2...i_{\frac{q}{2}}j_1j_2...j_{\frac{q}{2}}}(t)$ are independent Brownian variables with
\begin{equation}
\overline{J_{i_1i_2...i_{\frac{q}{2}}j_1j_2...j_{\frac{q}{2}}}(t)}=0,\ \ \ \ \ \ 
\overline{J_{i_1i_2...i_{\frac{q}{2}}j_1j_2...j_{\frac{q}{2}}}(t)J_{i_1i_2...i_{\frac{q}{2}}j_1j_2...j_{\frac{q}{2}}}(t')^*}=\frac{J\delta(t-t')}{\frac{q}{2}!(\frac{q}{2}-1)!N^{q-1}}.
\end{equation}
For the main part of this paper, we assume $q\geq 4$ and the system is many-body chaotic.

The time-dependent Hamiltonian \eqref{eqn:cBSYKH} has no energy conservation. Consequently, its steady states are at infinite temperature. Taking $U(1)$ charge conservation into account, the steady-state density matrix reads $\rho=\frac{1}{\mathcal{Z}}e^{-\mu Q}$ in the grand canonical ensemble, with total charge $Q=\sum_i c^\dagger_i c^{}_i$. The density of fermions is related to $\mu$ by
\begin{equation}
n=\frac{\left<Q\right>}{N}=\frac{1}{e^{\mu}+1}\in [0,1].
\end{equation}

We employ the Keldysh path-integral approach with partition function $\mathcal{Z}=\text{tr}\left(U(T)\rho U (T)^\dagger\right)$ and $U(T)=\mathcal{T}e^{-\int_{-T/2}^{T/2} H(t') dt'}$. The path-integral contour includes two branches $(u,d)$, which corresponds to the forward/backward evolution $(U,U^\dagger)$. A pictorial representation for $T\rightarrow \infty$ reads
\begin{equation}\label{eqn:singlecontoursketch}
\begin{tikzpicture}[scale=0.65,baseline={([yshift=0pt]current bounding box.center)}]

\draw[thick,blue,far arrow] (0pt,0pt)--(140pt,0pt);
\draw[thick,blue,far arrow] (140pt,-4pt)--(0pt,-4pt);
\filldraw (140pt,-2pt) circle (2pt) node[above right]{\scriptsize $\infty$};

\node at (100pt,10pt) {\scriptsize $u$};
\node at (100pt,-14pt) {\scriptsize $d$};

\filldraw (0pt,-2pt) circle (0pt) node[above left]{\scriptsize $-\infty$};

\draw[thick,blue] (0pt,0pt)--(0pt,10pt);
\draw[thick,blue] (-2pt,7pt)--(2pt,7pt);
\draw[thick,blue] (0pt,-4pt)--(0pt,-30pt);
\draw[thick,blue] (-2pt,-27pt)--(2pt,-27pt);

\node[svertex] at (0pt,-16pt) {\scriptsize$\rho$};

\end{tikzpicture}
\end{equation}
 The fermion fields $\psi_{i}^s$ and $\bar{\psi}_{i}^s$ on the Keldysh contour are labeled by branch indices $s\in \{u,d\}$. We first consider the fermion two-point functions $G^{ss'}(t)=\langle\psi^s_{i}(t)\bar{\psi}^{s'}_{i}(0)\rangle$. In the UV limit $t\rightarrow 0^{\pm}$, $G^{ss'}(t)$ can be determined by the density as 
 \begin{equation}
 G^{ud}(0)=-n,\ \ \ \ \ \ G^{du}(0)=1-n,\ \ \ \ \ \ G^{uu}(0^{\pm})=G^{dd}(0^{\mp})=\frac{1}{2}-n\pm\frac{1}{2}.
 \end{equation}
 
 To the leading order in $1/N$, the self-energy receives contributions from melon diagrams. The Schwinger-Dyson equation then reads
 \begin{equation}
 \begin{aligned}
 &\begin{pmatrix}
 \partial_t-\Sigma^{uu}&-\Sigma^{ud}\\
 -\Sigma^{du}& -\partial_t-\Sigma^{dd}
 \end{pmatrix}\circ
  \begin{pmatrix}
 G^{uu}&G^{ud}\\
 G^{du}& G^{dd}
 \end{pmatrix}=I,\ \ \ \ \ \ \Sigma^{ss'}=\begin{tikzpicture}[baseline={([yshift=-6pt]current bounding box.center)}, scale=1.2]
\draw[thick,mid arrow] (-24pt,0pt) -- (-15pt,0pt);
\draw[thick,dashed] (-15pt,0pt)..controls (-8pt,16pt) and (8pt,16pt)..(15pt,0pt);
\draw[thick,mid arrow] (-15pt,0pt)..controls (-8pt,10pt) and (8pt,10pt)..(15pt,0pt);
\draw[thick,mid arrow] (-15pt,0pt)..controls (-8pt,-10pt) and (8pt,-10pt)..(15pt,0pt);
\draw[thick,mid arrow] (15pt,0pt) -- (-15pt,0pt);
\draw[thick,mid arrow] (15pt,0pt) -- (24pt,0pt);
\end{tikzpicture}.
\end{aligned}
 \end{equation}
 Working out the details, we find 
  \begin{equation}
 \begin{aligned}
&\Sigma^{uu}(t)=-J\delta(t)G^{uu}(t)^{\frac{q}{2}}(-G^{uu}(-t))^{\frac{q}{2}-1}=-\frac{\Gamma}{2}(1-2n)\delta(t)=\Sigma^{dd}(t),\\
&\Sigma^{ud}(t)=-J\delta(t)G^{ud}(t)^{\frac{q}{2}}(-G^{du}(-t))^{\frac{q}{2}-1}=-\Gamma n\delta(t),\\
&\Sigma^{du}(t)=J\delta(t)G^{du}(t)^{\frac{q}{2}}(-G^{ud}(-t))^{\frac{q}{2}-1}=\Gamma (1-n)\delta(t),
\end{aligned}
 \end{equation}
 with the decay rate of quasi-particles $\Gamma=J \left(n(1-n)\right)^{\frac{q}{2}-1}$ \cite{chen2020many}. This gives
 \begin{equation}\label{eqn:BrownianG}
G(t)=\begin{pmatrix}
\frac{1}{2}-n+\frac{1}{2}\text{sgn}(t)&-n\\
1-n& \frac{1}{2}-n-\frac{1}{2}\text{sgn}(t)
\end{pmatrix}e^{-\frac{\Gamma |t|}{2}}.
 \end{equation}
Comparing to results in \cite{chen2020many}, \eqref{eqn:BrownianG} contains no running phase $e^{i\mu t}$, since our evolution Hamiltonian \eqref{eqn:cBSYKH} does not include the chemical potential term. It is also useful to work out the retarded and advanced Green's function as
\begin{equation}\label{eqn:GRGA}
G^\text{R}(t)=-i\theta(t)\text{tr}\left[\rho \{c_i^{}(t),c_i^{\dagger}(0)\}\right]=-i\theta(t)(G^{du}(t)-G^{ud}(t))=-i\theta(t)e^{-\Gamma t/2}=G^\text{A}(-t)^*.
\end{equation}
The dependence of density is only from the decay rate $\Gamma$. The result indicates the spectral function $A(\omega)=-2\text{Im}G^\text{R}(\omega)$ is Lorentzian, with a single peak near $\omega =0$.

\subsection{Wightman function with sources}\label{sec:wightman}
Now we turn to the study of OTO-correlations. Our final goal is to derive analytical expressions for the out-of-time-order correlators:
\begin{equation}\label{eqn:OTOCdef}
\begin{aligned}
&\text{OTOC}_1(t_1,t_2;t_3,t_4)=-\frac{1}{N^2}\sum_{ij}\text{tr}\left[\sqrt{\rho}c_i^{}(t_1)c_j^{}(t_3)\sqrt{\rho}c_i^\dagger(t_2)c_j^\dagger(t_4)\right],\\
&\text{OTOC}_2(t_1,t_2;t_3,t_4)=-\frac{1}{N^2}\sum_{ij}\text{tr}\left[\sqrt{\rho}c_i^{\dagger}(t_1)c_j^{}(t_3)\sqrt{\rho}c_i^{}(t_2)c_j^\dagger(t_4)\right].
\end{aligned}
\end{equation}
with $t_1 \approx t_2\gg t_3 \approx t_4$. Here we choose the convention to equally split the density matrix $\rho$, while results with other conventions can be computed straightforwardly using our results since $\rho$ commutes with the Hamiltonian. The OTOC can be understood as probing the perturbation distribution excited by operators in the past ($t_3,t_4$) using operators in the future ($t_1,t_2$) \cite{aleiner2016microscopic}. A trick proposed in \cite{gu2022two} is to replace the actual source with a mean-field perturbation and solve the Wightman function to determine the perturbation generated by a pair of operators, which gives an effective theory of scramblons. OTOCs are then given by combining two Wightman functions. 

We first study non-linear equations of Wightman functions on the double Keldysh contour in this subsection. Compared to the traditional Keldysh contour \eqref{eqn:singlecontoursketch}, it contains four branches, including two forward evolutions and two backward evolutions. We introduce a "world" label $w\in\{1,2\}$ in additional to $s\in\{u,d\}$, as indicated in the pictorial representation: 
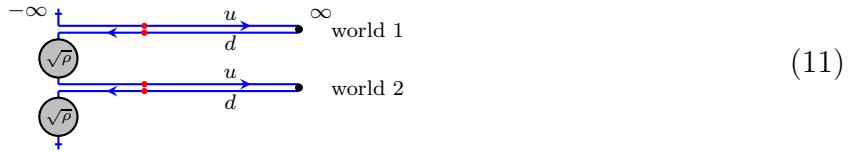
\begin{equation}\label{eqn:doublecontoursketch}
\begin{tikzpicture}[scale=0.65,baseline={([yshift=0pt]current bounding box.center)}]

\draw[thick,blue,far arrow] (0pt,0pt)--(140pt,0pt);
\draw[thick,blue,far arrow] (140pt,-4pt)--(0pt,-4pt);
\filldraw (140pt,-2pt) circle (2pt) node[above right]{\scriptsize $\infty$};

\node at (100pt,6pt) {\scriptsize $u$};
\node at (100pt,-10pt) {\scriptsize $d$};

\filldraw (0pt,-2pt) circle (0pt) node[above left]{\scriptsize $-\infty$};

\draw[thick,blue] (0pt,0pt)--(0pt,10pt);
\draw[thick,blue] (-2pt,7pt)--(2pt,7pt);
\draw[thick,blue] (0pt,-4pt)--(0pt,-34pt);

\node[svertex] at (0pt,-19pt) {\tiny$\sqrt{\rho}$};

\draw[thick,blue,far arrow] (0pt,-34pt)--(140pt,-34pt);
\draw[thick,blue,far arrow] (140pt,-38pt)--(0pt,-38pt);
\filldraw (140pt,-36pt) circle (2pt) node[above right]{};
\node at (100pt,-28pt) {\scriptsize $u$};
\node at (100pt,-44pt) {\scriptsize $d$};

\draw[thick,blue] (0pt,-38pt)--(0pt,-72pt);
\node[svertex] at (0pt,-53pt) {\tiny$\sqrt{\rho}$};
\draw[thick,blue] (-2pt,-69pt)--(2pt,-69pt);

\node at (180pt,-2pt) {\scriptsize world $1$};
\node at (180pt,-36pt) {\scriptsize world $2$};

\filldraw[red] (50pt,0pt) circle (1.5pt);
\filldraw[red] (50pt,-4pt) circle (1.5pt);

\filldraw[red] (50pt,-34pt) circle (1.5pt);
\filldraw[red] (50pt,-38pt) circle (1.5pt);

\end{tikzpicture}
\end{equation}
We can introduce Green's functions $G^{ss'}_{ww'}(t)=\langle\psi^{sw}_{i}(t)\bar{\psi}^{s'w'}_{i}(0)\rangle$. For $w=w'$, $G^{ss'}_{ww}(t)$ match the single Keldysh result \eqref{eqn:BrownianG} due to the unitarity. For $w \neq w'$, the unitarity implies $G^{ss'}_{ww'}(t)$ is independent of $s$ and $s'$. Explicitly, we have 
\begin{equation}
G_{21}^{ss'}(t)\equiv G^{\text{W},0}_{21}(t)=\sqrt{n(1-n)}e^{-\frac{\Gamma |t|}{2}}\equiv G(t),\ \ \ \ \ \ G_{12}^{ss'}(t)\equiv G^{\text{W},0}_{12}(t)=-G(t).
\end{equation}

Now we add mean-field sources to probe the perturbation created by fermion operators. We choose a source that does not affect single-world observables, such as $G^{ss'}_{ww}(t)$:
\begin{equation}\label{eqn:source}
\begin{aligned}
\delta S=&s_2 \sum_i(\overline{\psi}_{i}^{u2}(t_0)-\overline{\psi}_{i}^{d2}(t_0))({\psi}_{i}^{u1}(t_0)-{\psi}_{i}^{d1}(t_0))\\&+s_1 \sum_i({\psi}_{i}^{u2}(t_0)-{\psi}_{i}^{d2}(t_0))(\overline{\psi}_{i}^{u1}(t_0)-\overline{\psi}_{i}^{d1}(t_0)).
\end{aligned}
\end{equation}  
With the source term, the Wightman Green's functions $G^\text{W}_{12/21}(t,t')$ show explicit dependence of two time variables, which satisfies the Schwinger-Dyson equation \cite{aleiner2016microscopic}
\begin{equation}
\begin{aligned}
&\int dt_3dt_4~G^{\text{R}}(t_{13})\Sigma^\text{W}_{12}(t_3,t_4)G^{\text{A}}(t_{42})=G_{12}^\text{W}(t_1,t_2),\\
&\int dt_3dt_4~G^{\text{R}}(t_{13})\Sigma^\text{W}_{21}(t_3,t_4)G^{\text{A}}(t_{42})=G_{21}^\text{W}(t_1,t_2).
\end{aligned}
\end{equation} 
Here we have introduced $t_{ij}=t_i-t_j$ for conciseness. The self-energies receive contributions from source terms:
\begin{equation}
\begin{aligned}
\Sigma^\text{W}_{21}(t,t')&=J\delta(t-t')G^\text{W}_{21}(t,t)^{\frac{q}{2}}(-G^\text{W}_{12}(t,t))^{\frac{q}{2}-1}-s_2\delta(t-t_0)\delta(t'-t_0),\\
\Sigma^\text{W}_{12}(t,t')&=J\delta(t-t')G^\text{W}_{12}(t,t)^{\frac{q}{2}}(-G^\text{W}_{21}(t,t))^{\frac{q}{2}-1}+s_1\delta(t-t_0)\delta(t'-t_0).
\end{aligned}
\end{equation}
Using the explicit form in \eqref{eqn:GRGA}, we can inverse the retarded and advanced Green's functions and obtain differential equations
\begin{equation}\label{eqn:diffeqn}
\begin{aligned}
\left(\partial_{t_1}+\frac{\Gamma}{2}\right)\left(\partial_{t_2}+\frac{\Gamma}{2}\right)G_{w\bar{w}}^{\text{W}}(t_1,t_2)=\Sigma_{w\bar{w}}^{\text{W}}(t_1,t_2).
\end{aligned}
\end{equation}
Here we have introduced $\bar{w}\neq w$. The right-hand side of these equations only contains delta functions, which separate the solution into different regions as in the Majorana Brownian SYK case \cite{gu2022two}. Since the equations (and initial conditions) are symmetric with respect to $t_1$ and $t_2$, we assume
\begin{equation}
\left\{ 
\begin{aligned}
&G_{w\bar{w}}^\text{W}=e^{-\frac{\Gamma}{2}t_1}f_{w\bar{w}}(t_2)+e^{-\frac{\Gamma}{2}t_2}g_{w\bar{w}}(t_1), \ \ &(t_1,t_2)\in A,\\
&G_{w\bar{w}}^\text{W}=e^{-\frac{\Gamma}{2}t_2}f_{w\bar{w}}(t_1)+e^{-\frac{\Gamma}{2}t_1}g_{w\bar{w}}(t_2), \ \ &(t_1,t_2)\in B,\\
&G_{w\bar{w}}^{\text{W}}=(-1)^w\sqrt{n(1-n)}e^{-\frac{\Gamma}{2}|t_{12}|}, \ &(t_1,t_2)\in C\cup D,\\
&G_{w\bar{w}}^{\text{W}}(t_0^+,t_0^+)=G_{w\bar{w}}^{\text{W}}(t_0^-,t_0^-)-(-1)^ws_w.
\end{aligned}
\right.
\ \ \ \ \ \ \ \ \ 
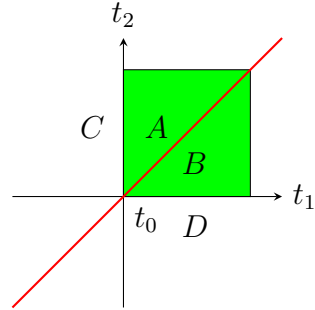
\begin{tikzpicture}[scale=0.6, baseline={(current bounding box.center)}]
\draw[->,>=stealth] (-70pt,0pt) -- (100pt,0pt) node[right]{$t_1$};
\draw[->,>=stealth] (0pt,-70pt) -- (00pt,100pt) node[above]{$t_2$};
\filldraw[fill=green] (0pt,0pt) rectangle (80pt,80pt);
\draw[red,thick] (-70pt,-70pt) -- (100pt,100pt);
\node at (35pt,30pt) [above left]{$A$};
\node at (30pt,35pt) [below right]{$B$};
\node at (-5pt,30pt) [above left]{$C$};
\node at (30pt,-5pt) [below right]{$D$};
\node at (0pt,0pt) [below right]{$t_0$};
\end{tikzpicture}
\end{equation}
For the third line, we use the fact that the source \eqref{eqn:source} can be neglected due to the cancellation between $u$ and $d$ branches for either $t_1<t_0$ or $t_2<t_0$. For the fourth line, we integrate the equation \eqref{eqn:diffeqn} over a small square surrounding $(t_0,t_0)$ and use the continuum condition without the source term $G_{w\bar{w}}^{\text{W}}(t_0^+,t_0^-)=G_{w\bar{w}}^{\text{W}}(t_0^-,t_0^+)=G_{w\bar{w}}^{\text{W}}(t_0^-,t_0^-)$. 

To determine the solution of $f_{w\bar{w}}$ and $g_{w\bar{w}}$, we match the boundary condition near the $AC$ and $AB$ boundary. Near the $AC$ boundary we have
\begin{equation}
(\partial_{t_2}+\frac{\Gamma}{2})f_{w\bar{w}}(t_2)=0,\ \ \ \ \ \ f_{w\bar{w}}=ae^{-\frac{\Gamma t_2}{2}}.
\end{equation}
We could always fix $a=0$ using the redundancy of $(f_{w\bar{w}}(t),g_{w\bar{w}}(t))\rightarrow (f_{w\bar{w}}(t)-ae^{-\frac{\Gamma t}{2}},g_{w\bar{w}}(t)+ae^{-\frac{\Gamma t}{2}})$. The boundary condition near $AB$ gives
\begin{equation}\label{eqn:diffwight}
\left(\frac{d}{dt}+\Gamma\right)z_{w\bar{w}}=\Gamma z_{w\bar{w}}^{\frac{q}{2}}z_{\bar{w}w}^{\frac{q}{2}-1},\ \ \ \ \ \ g_{w\bar{w}}(t)=(-1)^w\sqrt{n(1-n)}e^{\frac{\Gamma}{2} t}z_{w\bar{w}}(t).
\end{equation}
The solution can be parametrized as
\begin{equation}
z_{12}=\frac{C}{(1+ze^{\varkappa(t-t_0)})^{\frac{1}{q-2}}},\ \ \ \ \ \ z_{21}=\frac{C^{-1}}{(1+ze^{\varkappa(t-t_0)})^{\frac{1}{q-2}}}.
\end{equation}
Here $\varkappa=(q-2)\Gamma$ is the quantum Lyapunov exponent \cite{chen2020many}. The initial condition gives
\begin{equation}
1-\frac{s_2}{\sqrt{n(1-n)}}=\frac{C^{-1}}{(1+z)^{\frac{1}{q-2}}},\ \ \ \ \ \ 1-\frac{s_1}{\sqrt{n(1-n)}}=\frac{C}{(1+z)^{\frac{1}{q-2}}}.
\end{equation}
For small $s_1$ and $s_2$, we expand $\delta C=C-1$ and $z$ as
\begin{equation}
\delta C=\frac{s_2-s_1}{\sqrt{n(1-n)}},\ \ \ \ \ \ z=\frac{q-2}{2}\frac{s_2+s_1}{\sqrt{n(1-n)}}.
\end{equation}

We are interested in probing the distribution of scramblons perturbations. As a result, the trick is to take $s_i= p_ie^{\varkappa t_0}$ with $t_0 \rightarrow -\infty$. The final result is
\begin{equation}\label{eqn:GWres}
G^{\text{W}}_{w\bar{w}}(t_1,t_2)=\frac{G^{\text{W},0}_{w\bar{w}}(t_{12})}{(1+\tilde{z}e^{\varkappa\frac{t_1+t_2-|t_{12}|}{2}})^{\frac{1}{q-2}}},\ \ \ \ \ \ \tilde{z}=\frac{q-2}{2}\frac{p_2+p_1}{\sqrt{n(1-n)}}.
\end{equation}
Note that the result is symmetric under $p_1\leftrightarrow p_2$ and $w\leftrightarrow \bar{w}$, which suggests the OTO-correlations are symmetric for particles and holes. This leads to $\text{OTOC}_1=\text{OTOC}_2=\text{OTOC}$ in \eqref{eqn:OTOCdef}. As a result, one may set one of $p_i$ to be zero (we set $p_1=0$ and $p_2=p$), and derive scramblon data using a single source term. This will be explained in the next subsection.

\subsection{Scramblon diagrams and OTOC}
In the late-time regime, information scrambling is mediated by collective modes named scramblons \cite{kitaev2018soft}. In this subsection, we utilize results derived in the last subsection to extract the effective theory of scramblons in the cBSYK model. The OTOC can then be obtained by computing scramblon diagrams. 

We begin by examining $G^{\text{W}}$ in the effective theory of scramblons. The result \eqref{eqn:GWres} is derived to the leading order of $1/N$ with $-\log p_2\ll \varkappa t\ll\log N$, which can be identified with diagrams:
\begin{equation}\label{eqn:GWscrameblon}
G^\text{W}_{21}(t_1,t_2)=\begin{tikzpicture}
\node[bvertex] (R) at (-0pt,0pt) {};
\node[svertex] (A1) at (30/1.4142pt,30/1.4142pt) {\scriptsize $p$};
\node[svertex] (A2) at (30pt,0pt) {\scriptsize $p$};
\node[svertex] (A3) at (30/1.4142pt,-30/1.4142pt) {\scriptsize $p$};

\draw[thick] (R) -- ++(135:26pt) node[left]{$t_1$};
\draw[thick] (R) -- ++(-135:26pt) node[left]{$t_2$};
\draw[thick] (A1) -- ++(45+45:10pt) node{};
\draw[thick] (A1) -- ++(-45+45:10pt) node{};
\draw[thick] (A2) -- ++(45:10pt) node{};
\draw[thick] (A2) -- ++(-45:10pt) node{};
\draw[thick] (A3) -- ++(45-45:10pt) node{};
\draw[thick] (A3) -- ++(-45-45:10pt) node{};

\draw[wavy] (R) to (A1);
\draw[wavy] (R) to (A2);
\draw[wavy] (R) to (A3);
\end{tikzpicture}=\sum_{n}\frac{1}{n!}\left(\frac{-2pNe^{\varkappa \frac{t_1+t_2}{2} }\Upsilon^{\text{A},1}(0)}{C}\right)^n\Upsilon^{\text{R},n}(t_{12}).
\end{equation}
Here each insertion of the source term \eqref{eqn:source} create a scramblon, giving rise to a propagator $-e^{\varkappa t}/C$ with $C\propto N$. The factor of $2$ comes from two possible OTOCs generated by the source term and the factor of $N$ is due to the summation over indices in \eqref{eqn:source}. $\Upsilon^{\text{R/A},n}(t)$ is the vertex function in the future/past between a pair of fermions and $n$ scramblons. Due to the time-reversal symmetry, we have $\Upsilon^{\text{R},n}(t)=\Upsilon^{\text{A},n}(t)$.

Noticing \eqref{eqn:GWscrameblon} is invariant under $C\rightarrow \lambda^2 C$ and $\Upsilon^{\text{R/A},n} \rightarrow \lambda^n \Upsilon^{\text{R/A},n}$, which reflects an arbitrary rescaling for the definition of scramblons. For simplicity, we fix the convention that
\begin{equation}\label{eqn:convention}
\frac{2N}{C}\Upsilon^{\text{A},1}(0)=\frac{q-2}{2\sqrt{n(1-n)}},
\end{equation}
which leads to
\begin{equation}\label{eqn:resf}
f(x,t)=\sum_{n!}\frac{(-x)^n}{n!}\Upsilon^{\text{R},n}(t)=\frac{G(t)}{(1+xe^{-\frac{\varkappa}{2}|t|})^{\frac{1}{q-2}}}.
\end{equation}
Performing the Taylor expansion, we find
\begin{equation}
\Upsilon^{\text{R},n}(t)=G(t)\frac{\Gamma(\frac{1}{q-2}+n)}{\Gamma(\frac{1}{q-2})}e^{-\frac{n\varkappa}{2}|t|}=\Upsilon^{\text{A},n}(t).
\end{equation}
Together with the convention \eqref{eqn:convention}, this fixes
\begin{equation}
\Upsilon^{\text{R},1}(0)=\Upsilon^{\text{A},1}(0)=\frac{\sqrt{n(1-n)}}{q-2},\ \ \ \ \ \ C=\frac{4Nn(1-n)}{(q-2)^2}.
\end{equation}

The late-time OTOC is defined for the time regime with $\varkappa t\sim \log N$. To the leading order in $1/N$ expansion, we have 
\begin{equation}
\begin{aligned}
\text{OTOC}(t_1,t_2;t_3,t_4)=&\begin{tikzpicture}
\node[bvertex] (R) at (-30pt,0pt) {};
\node[bvertex] (A) at (30pt,0pt) {};
\draw[thick] (R) -- ++(135:20pt) node[left]{$t_1$};
\draw[thick] (R) -- ++(-135:20pt) node[left]{$t_2$};
\draw[thick] (A) -- ++(45:20pt) node[right]{$t_3$};
\draw[thick] (A) -- ++(-45:20pt) node[right]{$t_4$};
\draw[wavy] (A) to[out=140,in=40] (R);
\draw[wavy] (A) to (R);
\draw[wavy] (A) to[out=-140,in=-40] (R);
\end{tikzpicture}=\sum_n\left(-\lambda\right)^n\Upsilon^{\text{R},n}(t_{12})\Upsilon^{\text{A},n}(t_{34}).
\end{aligned}
\end{equation}
Since $f(x,t)$ is analytic expect a branch cut along the negative real axis, we could introduce its inverse Laplace transform $h(x,t)$ as in \cite{gu2022two}:
\begin{equation}
f(x,t)=\int dy~h(y,t)e^{-xy},\ \ \ \ \ \ \Upsilon^{\text{R},n}(t)=\int dy~h(y,t)y^n.
\end{equation}
Using \eqref{eqn:resf}, we have 
\begin{equation}
h(y,t)=\sqrt{n(1-n)}\frac{y^{\frac{1}{q-2}-1}}{\Gamma(\frac{1}{q-2})}e^{-ye^{\frac{\varkappa}{2}|t|}}.
\end{equation}
 This leads to several equivalent expressions for the OTOC:
\begin{equation}
\begin{aligned}
\text{OTOC}(t_1,t_2;t_3,t_4)&=\int dy_\text{A}dy_\text{R}~h(y_\text{A},t_{12})h(y_\text{R},t_{34})e^{-\lambda y_\text{A}y_\text{R}}\\
&=\int dy_\text{A}~h(y_\text{A},t_{12})f(\lambda y_\text{A},t_{34})=\int dy_\text{R}~f(\lambda y_\text{R},t_{12})h(y_\text{R},t_{34}).
\end{aligned}
\end{equation}
In this first line, the result can be understood as each pair of operators creates perturbations in the future/past with distribution $h(y_{\text{R/A}},t)$, which interact through an Euclidean action $S_{\text{eff}}=\lambda y_\text{A}y_\text{R}$. In the second line, we integrate out $y_{\text{R}}$ (or $y_{\text{A}}$), and the pair of operators in the past (future) serves as a probe of perturbations in the future (past), with a probe function $f(\lambda y_\text{A/R},t)$. Finally, performing the remaining integral explicitly, we find
\begin{equation}
\begin{aligned}
\text{OTOC}(t_1,t_2;t_3,t_4)=G(t_{12})G(t_{34})\left[\frac{e^{\frac\varkappa2 (|t_{12}|+|t_{34}|)}}{\lambda}\right]^{\frac{1}{q-2}}U\left(\frac{1}{q-2},1,\frac{e^{\frac\varkappa2 (|t_{12}|+|t_{34}|)}}{\lambda}\right).
\end{aligned}
\end{equation} 
Here we have introduced $\lambda=\frac{e^{\varkappa\frac{t_1+t_2-t_3-t_4}{2}}}{C}$. The result shows that the charge density dependence of $\frac{\text{OTOC}(t_1,t_2;t_3,t_4)}{G(t_{12})G(t_{34})}$ only comes from $\lambda$, consistent with the numerical observations in \cite{agarwal2022emergent}. This extends previous discussions on information scrambling in the cBSYK model to the late-time regime \cite{chen2020many}.

\subsection{Late-time operator size distribution}
Finally, we consider the operator size distribution of the cBSYK model. The operator size is defined unambiguously at infinite temperature in the full Hilbert space. In this subsection, we first explain our proposal for extending the definition to systems with charge conservation in the grand canonical ensemble with $\mu \neq 0$, which is an analog of the finite-temperature size for systems with energy conservations \cite{sizenewpaper,qi2019quantum,lucas2019operator}. We then derive a concrete formula for the late-time operator size using scramblon diagrams. 

We begin with the $\mu=0$ case. To be concrete, we consider the Heisenberg evolution $c_i(t)=U(t)^\dagger c_i(0)U(t)$. The definition of the operator size is basis dependent. Here we choose the local orthonormal basis $\{O_j^a\}=\{I,c_j+c_j^\dagger,i(c_j-c_j^\dagger) , 2c_j^\dagger c_j-1\}$, and define their operator size as $\{n^a\}=\{0,1,1,2\}$. The operator basis for the total system is given by a tensor product of local basis $O_1^{a_1}O_2^{a_2}...O_N^{a_N}$ with an operator size $S=\sum_jn^{a_1}_j$.  This definition matches the convention for Majorana fermions \cite{roberts2018operator}. The operator size distribution $P(S)$ of $c_i(t)$ is then defined by 
\begin{equation}
c_i(t)=\sum_{\{a_n\}}c_{a_1a_2...a_N}O_1^{a_1}O_2^{a_2}...O_N^{a_N},\ \ \ \ \ \ P(S)=2\sum_{\{a_n\}}|c_{a_1a_2...a_N}|^2\delta_{\sum_jn^{a_1}_j,S}.
\end{equation}
At late-time regime, we take the continuum limit of the operator size by introducing $s\equiv S/N \in[0,2]$ and $\mathcal{P}(s)\equiv NP(sN)$. It is straightforward to show $\int ds~\mathcal{P}(s)=\sum_S P(S)=2\langle c^\dagger_jc_j\rangle=1$. 

To compute the operator size distribution, we use the trick by introducing an auxiliary system with $N$ complex fermions $\xi_j$. We first prepare the initial state as a maximally entangled state between $c_j$ and $\xi_j$. On the occupation basis, we have
\begin{equation}
|I\rangle=\prod_{ j}\otimes\frac{1}{\sqrt{2}}(|01\rangle_j+|10\rangle_j).
\end{equation}
The state satisfies $(c_j-\xi_j)|I\rangle=(c_j^\dagger+\xi_j^\dagger)|I\rangle=0$. This motivates us to measure the perturbation of $|I\rangle$ using operator 
\begin{equation}
Q_S=\frac{1}{2}\sum_j\left[(c_j^\dagger-\xi_j^\dagger)(c_j^{}-\xi_j^{})+(c_j^{}+\xi_j^{})(c_j^\dagger+\xi_j^{\dagger})\right].
\end{equation}
It is straightforward to show that
\begin{equation}
Q_S|I\rangle=0,\ \ \ Q_Sc^{}_j|I\rangle=c^{}_j|I\rangle,\ \ \ Q_Sc^{\dagger}_j|I\rangle=c^{\dagger}_j|I\rangle,\ \ \ Q_S(2c^{\dagger}_jc^{}_j-1)|I\rangle=2(2c^{\dagger}_jc^{}_j-1)|I\rangle.
\end{equation}
This shows that the eigenvalue of $Q_S$ matches the definition of operator size. As a result, the generating function of the operator size distribution can be expressed as
\begin{equation}
\mathcal{S}(\nu)=\int ds~\mathcal{P}(s)e^{-\nu s}=2\langle I|c^\dagger_j(t)e^{-\frac{\nu Q_S}{N}}c^{}_j(t)|I\rangle.
\end{equation} 

Now we generalize above arguments to finite $\mu$. We first consider applying $\sqrt{\rho_c}\equiv \sqrt{\rho}\otimes I$ to $|I\rangle$, which leads to a bias in the Hilbert space
\begin{equation}
|I_\mu\rangle=\sqrt{\rho_c}|I\rangle=\frac{1}{\sqrt{\mathcal{Z}}}\prod_{ j}\otimes\frac{1}{\sqrt{2}}(|01\rangle_j+e^{-\frac{\mu}{2}}|10\rangle_j).
\end{equation}
Tracing out the auxiliary system leads to the correction density matrix of $\text{tr}_\xi(|I_\mu\rangle\langle I_\mu|)=\rho$. To probe deviations from $|I_\mu\rangle$, we ask which operator annihilates the state. We take a symmetric convention with
\begin{equation}
(e^{\frac{\mu}{4}}c_j-e^{-\frac{\mu}{4}}\xi_j)|I_\mu\rangle=(\rho_c^{\frac{1}{4}}c_j\rho_c^{-\frac{1}{4}}-\rho_\xi^{\frac{1}{4}}\xi_j\rho_\xi^{-\frac{1}{4}})|I_\mu\rangle=\rho_c^{\frac{1}{4}}\rho_\xi^{\frac{1}{4}}(c_j-\xi_j)|I\rangle=0
\end{equation}
Here we have introduced $\rho_\xi \equiv I \otimes {e^{\mu \sum_j \xi_j^\dagger\xi_j}}/\mathcal{Z}$ and used $\rho_\xi|I\rangle= \rho_c |I\rangle$. Similarly, we have $(e^{-\frac{\mu}{4}}c_j^\dagger+e^{\frac{\mu}{4}}\xi_j^\dagger)|I_\mu\rangle=0$. As in the $\mu=0$ case, we introduce the positive semi-definite operator 
\begin{equation}
Q_{S\mu}\equiv \frac{\cosh{\frac{\mu}{2}}}{2}\sum_j\left[(e^{\frac{\mu}{4}}c_j^\dagger-e^{-\frac{\mu}{4}}\xi_j^\dagger)(e^{\frac{\mu}{4}}c_j^{}-e^{-\frac{\mu}{4}}\xi_j^{})+(e^{-\frac{\mu}{4}}c_j^{}+e^{\frac{\mu}{4}}\xi_j^{})(e^{-\frac{\mu}{4}}c_j^\dagger+e^{\frac{\mu}{4}}\xi_j^{\dagger})\right],
\end{equation}
and define the finite density operator size of $c_j(t)$ by the generating function 
\begin{equation}
\begin{aligned}
\mathcal{S}_\mu(\nu)&=\int ds~\mathcal{P}_\mu(s)e^{-\nu s}\equiv\frac{\langle I|c^\dagger_j(t)\rho_c^{\frac{1}{4}}\rho_\xi^{\frac{1}{4}}e^{-\frac{\nu Q_{S\mu}}{N}}\rho_c^{\frac{1}{4}}\rho_\xi^{\frac{1}{4}}c^{}_j(t)|I\rangle}{\langle I|c^\dagger_j(t)\rho_c^{\frac{1}{2}}\rho_\xi^{\frac{1}{2}}c^{}_j(t)|I\rangle}\\&=2\cosh\frac{\mu}{2}\langle I|c^\dagger_j(t)\rho_c^{\frac{1}{4}}\rho_\xi^{\frac{1}{4}}e^{-\frac{\nu Q_{S\mu}}{N}}\rho_c^{\frac{1}{4}}\rho_\xi^{\frac{1}{4}}c^{}_j(t)|I\rangle.
\end{aligned}
\end{equation} 
In the late-time regime with $\varkappa t\sim \log N$, only contributions from scramblon diagrams are important due to the suppression of $1/N$. As in \cite{sizenewpaper}, the result can be derived by arguments similar to that of the OTOC: We imagine the insertion of $c_j$ and $c_j^\dagger$ creates the perturbation in the future, whose strength is described by the distribution function $h^\text{R}(y,0)$. This perturbation is probed by the size operator $Q_{S\mu}$ is the past, which has a probe function
\begin{equation}
Q_{S\mu}\approx N\left(1-2 \cosh\left(\frac{\mu}{2}\right)f(\lambda y,0)\right)=N\left(1-\frac{1}{(1+\lambda y)^{\frac{1}{q-2}}}\right),\ \ \ \ \ \ \lambda=e^{\varkappa t}/C.
\end{equation}
Here we take the expectation value over $|I_\mu\rangle$ for terms without OTO-correlations. This leads to the result
\begin{equation}
\mathcal{S}_\mu(\nu)=2\cosh\frac{\mu}{2}\int dy~h(y,0)e^{-\nu [1-{(1+\lambda y)^{-\frac{1}{q-2}}}] }=\int dy~\frac{y^{\frac{1}{q-2}-1}}{\Gamma(\frac{1}{q-2})}e^{-y}e^{-\nu [1-{(1+\lambda y)^{-\frac{1}{q-2}}}] }.
\end{equation}
Performing the inverse Laplace transform, the distribution $\mathcal{P}(s)$ is given by
\begin{equation}
\mathcal{P}_\mu(s)=\left|y'(s)\right|\frac{y(s)^{\frac{1}{q-2}-1}}{\Gamma(\frac{1}{q-2})}, \ \ \ \ \ \ s=1-{(1+\lambda y)^{-\frac{1}{q-2}}}\in[0,1].
\end{equation}
In the early-time regime with $\lambda\ll 1$, the operator size grows exponentially with exponent $\varkappa$. In the long-time limit, $c_j(t)$ approaches a maximally scrambled operator, which leads to $\mathcal{P}(s)=\delta(s-1)$, sinece the standard deviation of the operator size can be neglected to the leading order of $1/N$. Moreover, as for OTOC, the density dependence only comes from the propagator $\lambda$ of scramblons \footnote{This statement depends on the overall coefficient in the definition of $Q_{S\mu}$. One may change the definition by an overall factor, which leads to a rescale of $s$. }. This generalizes previous results for the late-time operator size distribution of Majorana SYK models \cite{sizenewpaper}.

\section{Entanglement dynamics and transitions}

In this section, we consider the entanglement dynamics of complex Brownian SYK chains. We first consider the unitary evolutions and study the charge dependence of the entanglement velocity for both the Von Neumann entropy and the $m$-th R\'enyi entropy. We then add repeated weak measurements, where measurement induced phase transitions\cite{Li_2018,Cao_Tilloy_2019,Li_2019,Skinner_2019,Chan_2019,Bao_2020,Choi_2020,gullans2019dynamical,gullans2019scalable,zabalo2019critical,Tang_Zhu_2020,Szyniszewski_2019,Zhang_2020,biella2021many} exist for both interacting and non-interacting models. We derive the effective theory for the transition, with a comparison to its Majorana counterparts \cite{liu2021non,zhang2021emergent,jian2021measurement,zhang2022quantum}.

\subsection{Model and set-up}\label{entanglementsetup}
We extend the cBSYK model \eqref{eqn:cBSYKH} to 1-D chains by introducing multiple copies and adding Brownian hopping terms \cite{jian2021note}. The Hamiltonian reads
\begin{equation}\label{eqn:cBSYKH_chain}
\begin{aligned}
H(t)=&\sum_x\sum_{i_1<i_2...<i_{\frac{q}{2}}}\sum_{j_1<j_2...<j_{\frac{q}{2}}}J_{i_1i_2...i_{\frac{q}{2}}j_1j_2...j_{\frac{q}{2}}}^x(t)c^\dagger_{i_1,x}c^\dagger_{i_2,x}...c^\dagger_{i_{\frac{q}{2}},x}c^{}_{j_1,x}c^{}_{j_2,x}...c^{}_{j_{\frac{q}{2}},x}\\
&+\sum_{x}\left[\sum_{ij}V_{ij}^x(t)c^\dagger_{i,x+1}c^{}_{j,x}+\text{H.C.}\right].
\end{aligned}
\end{equation}
Here we take the periodic boundary condition with $x\in\{1,2,...,L\}$. Random parameters on different sites are labeled by $x$ and thus independent. The random hopping strength $V_{ij}^x(t)$ are Brownian variables with
\begin{equation}
\overline{V_{ij}^x(t)}=0,\ \ \ \ \ \ \overline{V_{ij}^x(t)V_{ij}^x(t')^*}=\frac{V}{2N}\delta(t-t').
\end{equation}
Following the the discussion in section \ref{cBSYKsub1}, we find the Green's functions on the Keldysh contour $G_x^{ss'}(t)=\langle\psi^s_{i}(t)\bar{\psi}^{s'}_{i}(0)\rangle$ are still given by \eqref{eqn:BrownianG}, with decay rate $\Gamma=J \left(n(1-n)\right)^{\frac{q}{2}-1}+V$. 

In this section, we are interested in the dynamics of the entanglement entropy for pure states. We focus on the setup with an auxiliary fermion chain $\xi_{i,x}$ \cite{jian2021note,chen2020replica}, which is a direct analog of the gravity calculation \cite{almheiri2019islands}. The system is prepared in 
\begin{equation}
|\psi_0\rangle=\prod_{x} \otimes|I_\mu\rangle_{x},
\end{equation}
which contains no entanglement between different sites $x$. The system is then evolved under the Hamiltonian \eqref{eqn:cBSYKH_chain}, where the entanglement between different sites builds up. We then choose first $L_A$ sites $x\in [1,2,...,L_A]$ as the subsystem $A$ including both $c_{i,x}$ and $\xi_{i,x}$ fermions. The reduced density matrix $\rho_A$ reads
\begin{equation}
\begin{aligned}
\rho_A&=\text{tr}_{\bar{A}}~U(t)|\psi_0\rangle \langle\psi_0|U(t)^\dagger=\begin{tikzpicture}[scale=0.65,baseline={([yshift=-3pt]current bounding box.center)}]

\draw[thick,blue,far arrow] (0pt,0pt)--(70pt,0pt);

\filldraw (70pt,-2pt) circle (0pt) node[above right]{\scriptsize $c_{A}$};

\filldraw (35pt,-2pt) circle (0pt) node[above]{\scriptsize $u$};

\filldraw (0pt,-2pt) circle (0pt) node[above left]{\scriptsize $\xi_A$};

\draw[thick,red,far arrow] (0pt,-10pt)--(70pt,-10pt);
\draw[thick,red,far arrow] (70pt,-14pt)--(0pt,-14pt);
\filldraw (70pt,-12pt) circle (2pt) node[above right]{};
\filldraw (0pt,-12pt) circle (2pt) node[above right]{};

\draw[thick,blue,far arrow] (70pt,-24pt)--(0pt,-24pt);

\filldraw (70pt,-22pt) circle (0pt) node[below right]{\scriptsize $c_{A}$};
\filldraw (35pt,-22pt) circle (0pt) node[below]{\scriptsize $d$};

\filldraw (0pt,-22pt) circle (0pt) node[below left]{\scriptsize $\xi_A$};

\draw[thick,dotted] (14pt,-14pt)--(14pt,-24pt);
\draw[thick,dotted] (28pt,-14pt)--(28pt,-24pt);
\draw[thick,dotted] (42pt,-14pt)--(42pt,-24pt);
\draw[thick,dotted] (56pt,-14pt)--(56pt,-24pt);

\draw[thick,dotted] (14pt,-10pt)--(14pt,0pt);
\draw[thick,dotted] (28pt,-10pt)--(28pt,0pt);
\draw[thick,dotted] (42pt,-10pt)--(42pt,0pt);
\draw[thick,dotted] (56pt,-10pt)--(56pt,0pt);

\end{tikzpicture}
\end{aligned}
\end{equation}
Here we draw a pictorial representation, which is understood as a path-integral over the corresponding contour \cite{chen2020replica,zhang2020entanglement}. We have omitted the density matrix $\rho$ for simplicity. Dotted lines represent the interaction between A and $\bar{A}$ in the unitary evolution. We are interested in the entropy of $\rho_A$. The $m$-th R\'enyi entropy and the Von Neumann entropy are defined as
\begin{equation}
S^{(m)}_A=-\frac{1}{m-1}\log \text{tr}_A(\rho_A^n),\ \ \ \ \ \ S_A=\underset{m\rightarrow 1}{\lim}~S^{(m)}_A=-\text{tr}_A(\rho_A\log \rho_A).
\end{equation}
As an example, for the second and the forth R\'enyi entropy, we have 
\begin{equation}\label{eqn:contour}
\text{tr}_A(\rho_A^2)\ =\ 
\begin{tikzpicture}[scale=0.65,baseline={([yshift=-5pt]current bounding box.center)}]

\draw[thick,blue,far arrow] (-70pt,2pt)--(-10pt,2pt);
\draw[thick,blue,far arrow] (-10pt,-2pt)--(-70pt,-2pt);
\filldraw (-70pt,0pt) circle (2pt) node[left]{\scriptsize 1};
\filldraw (-10pt,0pt) circle (2pt) node[above right]{};

\draw[thick,blue,far arrow] (10pt,2pt)--(70pt,2pt);
\draw[thick,blue,far arrow] (70pt,-2pt)--(10pt,-2pt);
\filldraw (10pt,0pt) circle (2pt) node[right]{};
\filldraw (70pt,0pt) circle (2pt) node[right]{\scriptsize 2};

\draw[thick,red,far arrow] (2pt,-70pt)--(2pt,-10pt);
\draw[thick,red,far arrow] (-2pt,-10pt)--(-2pt,-70pt);
\filldraw (0pt,-70pt) circle (2pt) node[below]{\scriptsize 2};
\filldraw (0pt,-10pt) circle (2pt) node[above right]{};

\draw[thick,red,far arrow] (2pt,10pt)--(2pt,70pt);
\draw[thick,red,far arrow] (-2pt,70pt)--(-2pt,10pt);
\filldraw (0pt,70pt) circle (2pt) node[above]{\scriptsize 1};
\filldraw (0pt,10pt) circle (2pt) node[above right]{};

\draw[thick,dotted] (-50pt,2pt)--(-2pt,50pt);
\draw[thick,dotted] (-30pt,2pt)--(-2pt,30pt);

\draw[thick,dotted] (-50pt,-2pt)--(-2pt,-50pt);
\draw[thick,dotted] (-30pt,-2pt)--(-2pt,-30pt);

\draw[thick,dotted] (50pt,2pt)--(2pt,50pt);
\draw[thick,dotted] (30pt,2pt)--(2pt,30pt);

\draw[thick,dotted] (50pt,-2pt)--(2pt,-50pt);
\draw[thick,dotted] (30pt,-2pt)--(2pt,-30pt);

\end{tikzpicture}\ \ \ \ \ \ \text{tr}_A(\rho_A^4)\ =\ 
\begin{tikzpicture}[scale=0.65,baseline={([yshift=-5pt]current bounding box.center)}]

\draw[thick,blue,far arrow] (-70pt,2pt)--(-10pt,2pt);
\draw[thick,blue,far arrow] (-10pt,-2pt)--(-70pt,-2pt);
\filldraw (-70pt,0pt) circle (2pt) node[left]{\scriptsize 1};
\filldraw (-10pt,0pt) circle (2pt) node[above right]{};

\draw[thick,blue,far arrow] (10pt,2pt)--(70pt,2pt);
\draw[thick,blue,far arrow] (70pt,-2pt)--(10pt,-2pt);
\filldraw (10pt,0pt) circle (2pt) node[right]{};
\filldraw (70pt,0pt) circle (2pt) node[right]{\scriptsize 3};

\draw[thick,blue,far arrow] (2pt,-10pt)--(2pt,-70pt);
\draw[thick,blue,far arrow] (-2pt,-70pt)--(-2pt,-10pt);
\filldraw (0pt,-70pt) circle (2pt) node[below]{\scriptsize 4};
\filldraw (0pt,-10pt) circle (2pt) node[above right]{};

\draw[thick,blue,far arrow] (2pt,70pt)--(2pt,10pt);
\draw[thick,blue,far arrow] (-2pt,10pt)--(-2pt,70pt);
\filldraw (0pt,70pt) circle (2pt) node[above]{\scriptsize 2};
\filldraw (0pt,10pt) circle (2pt) node[above right]{};

\draw[thick,red,far arrow] (-5.65685pt,8.48528pt)--(-48.0833pt,50.9117pt);
\draw[thick,red,far arrow] (-50.9117pt,48.0833pt)--(-8.48528pt,5.65685pt);
\filldraw (-49.4975pt,49.4975pt) circle (2pt) node[above left]{\scriptsize 1};
\filldraw (-7.07107pt,7.07107pt) circle (2pt) node[above right]{};

\draw[thick,red,far arrow] (48.0833pt,50.9117pt)--(5.65685pt,8.48528pt);
\draw[thick,red,far arrow] (8.48528pt,5.65685pt)--(50.9117pt,48.0833pt);
\filldraw (49.4975pt,49.4975pt) circle (2pt) node[above right]{\scriptsize 2};
\filldraw (7.07107pt,7.07107pt) circle (2pt) node[above right]{};

\draw[thick,red,far arrow] (-48.0833pt,-50.9117pt)--(-5.65685pt,-8.48528pt);
\draw[thick,red,far arrow] (-8.48528pt,-5.65685pt)--(-50.9117pt,-48.0833pt);
\filldraw (-49.4975pt,-49.4975pt) circle (2pt) node[below left]{\scriptsize 4};
\filldraw (-7.07107pt,-7.07107pt) circle (2pt) node[above right]{};

\draw[thick,red,far arrow] (5.65685pt,-8.48528pt)--(48.0833pt,-50.9117pt);
\draw[thick,red,far arrow] (50.9117pt,-48.0833pt)--(8.48528pt,-5.65685pt);
\filldraw (49.4975pt,-49.4975pt) circle (2pt) node[below right]{\scriptsize 3};
\filldraw (7.07107pt,-7.07107pt) circle (2pt) node[above right]{};

\draw[thick,dotted] (-50pt,2pt)--(-36.7696pt,33.9411pt);
\draw[thick,dotted] (-30pt,2pt)--(-22.6274pt,19.799pt);

\draw[thick,dotted] (-50pt,-2pt)--(-36.7696pt,-33.9411pt);
\draw[thick,dotted] (-30pt,-2pt)--(-22.6274pt,-19.799pt);

\draw[thick,dotted] (50pt,2pt)--(36.7696pt,33.9411pt);
\draw[thick,dotted] (30pt,2pt)--(22.6274pt,19.799pt);

\draw[thick,dotted] (50pt,-2pt)--(36.7696pt,-33.9411pt);
\draw[thick,dotted] (30pt,-2pt)--(22.6274pt,-19.799pt);

\draw[thick,dotted] (2pt,-50pt)--(33.9411pt,-36.7696pt);
\draw[thick,dotted] (2pt,-30pt)--(19.799pt,-22.6274pt);

\draw[thick,dotted] (-2pt,-50pt)--(-33.9411pt,-36.7696pt);
\draw[thick,dotted] (-2pt,-30pt)--(-19.799pt,-22.6274pt);

\draw[thick,dotted] (2pt,50pt)--(33.9411pt,36.7696pt);
\draw[thick,dotted] (2pt,30pt)--(19.799pt,22.6274pt);

\draw[thick,dotted] (-2pt,50pt)--(-33.9411pt,36.7696pt);
\draw[thick,dotted] (-2pt,30pt)--(-19.799pt,22.6274pt);
\end{tikzpicture}
\end{equation}
Here we have world index $w\in\{1,2,...,m\}$. Using the standard SYK technique, to the leading order of $1/N$ the R\'enyi entropy is given by the $G$-$\Sigma$ action:
\begin{equation}\label{eqn:GSigma}
\begin{aligned}
(m-1)\frac{S^{(m)}_A}{N}&=\min_{G\Sigma}\sum_{x}\left\{-\text{tr}\log \left(\eta_s\delta_{ww'}^{ss'}\partial_t-\Sigma^{ss'}_{ww',x}\right)-\int dt~\Sigma^{ss'}_{ww',x}(t,t)G^{s's}_{w'w,x}(t,t)\right.\\
&\left.+\int dt~\eta_{s}\eta_{s'}\left[\frac{J}{q}(-G^{s's}_{w'w,x}G^{ss'}_{ww',x})^{\frac{q}{2}}-\frac{V}{2}T^{xs}_{w_1w_2}T^{xs'}_{w_3w_4}G^{s's}_{w_3w_1,x}G^{ss'}_{w_2w_4,x+1}\right]-C_0\right\}.
\end{aligned}
\end{equation}
Here we have introduced $\eta_u=1$, $\eta_d=-1$. $T^{xs}_{ww'}=\delta_{ww'}$ except a twist near the boundary between system $A$ and $\bar{A}$: $T^{L_A d}_{ww'}=T^{L d}_{w'w}=\delta_{w,w'+1}$. The constant $C_0$ is chosen such that $S_{\varnothing}^{(m)}=0$. In general, a full analytical study of \eqref{eqn:GSigma} is impossible, and numerical simulations for its saddle-point equations are employed. In this section, we instead focus on certain limits where analytical formula can be derived, as explained in the following subsections.

\subsection{Density dependence of entanglement velocity}
In this subsection, we are interested in the entanglement velocity, in particular its density dependence, of the cBSYK chain. The entanglement velocity $v_E^{(m)}$ is defined as the slope of entropy in the early-time regime $S_A^{(m)}\approx 2v_E^{(m)}s_0^{(m)} t$, where $s_0^{(m)}=-{N}(m-1)^{-1}\log(n^m+(1-n)^m)$ is the maximal entropy density\footnote{Here the factor of 2 is due to the exsitence of two boundaries in our setup.}. We focus on small hopping strength $V$, where $v_E^{(m)}$ can be obtained by a perturbative calculation. Our discussion primarily follows \cite{dadras2021perturbative}, which focuses on static system-bath couplings with Hermitian operators. 

We begin with the R\'enyi entropy with $n>1$. For $V=0$, the contour \eqref{eqn:contour} is reduced to $n$ copies of the traditional Keldysh contour \eqref{eqn:singlecontoursketch}. As a result, the Green's functions is diagonal in the world index $G^{ss'}_{ww',x}(t)=G^{ss'}(t)\delta_{ww'}$ with $G^{ss'}(t)$ given by \eqref{eqn:BrownianG}. Since there is no coupling between different sites, we have $S_A^{(m)}=0$ for $V=0$. Now we adding the effect of $V$ perturbatively. The leading order contribution comes from evaluating the $V$ term in \eqref{eqn:GSigma} using Green's functions with $V=0$. Due to the unitarity, the only contribution is from boundary terms with $x=L_A$ and $x=L$:
\begin{equation}\label{eqn:SmA}
S_A^{(m)}(t)=-\frac{mN}{m-1}V\int_0^t dt'~\sum_sG^{ss}(t',t')G^{ss}(t',t')=\frac{2mN}{m-1}Vn(1-n)t.
\end{equation}
More generally, in models where nearest neighbor sites are coupled through $p$-body Brownian term $\sim(c^{\dagger}_{x+1}c^{}_{x})^{p/2}$, we expect $v_E^{(m)}s_0^{(m)}\sim V [n(1-n)]^{\frac p2}$. 

It is also interesting to compare $v_E^{(m)}s_0^{(m)}$ with butterfly velocity $v_B$. To determine $v_B$, we generalize the discussion in subsection \ref{sec:wightman} to the SYK chain case. \eqref{eqn:diffwight} now becomes
\begin{equation}\label{eqn:GWequationchain}
\left(\frac{d}{dt}+\Gamma_J+\Gamma_V\right)z_{w\bar{w},x}=\Gamma_J z_{w\bar{w},x}^{\frac{q}{2}}z_{\bar{w}w,x}^{\frac{q}{2}-1}+\frac{\Gamma_V}{2}(z_{w\bar{w},x+1}^{\frac{p}{2}}z_{\bar{w}w,x}^{\frac{p}{2}-1}+z_{w\bar{w},x-1}^{\frac{p}{2}}z_{\bar{w}w,x}^{\frac{p}{2}-1}).
\end{equation}
Here $\Gamma_J=J \left(n(1-n)\right)^{\frac{q}{2}-1}$ and $\Gamma_V=V \left(n(1-n)\right)^{\frac{p}{2}-1}$. Now we consider the linearizion of above equation $z_{\bar{w}w,x}=1-\delta z_{\bar{w}w,x}$ with $z_{\bar{w}w,x}=z_{w\bar{w},x}$, which gives
\begin{equation}
\frac{d}{dt}\delta z_{w\bar{w},x}=(q-2)\Gamma_J\delta z_{w\bar{w},x}+\frac{p}{2}\frac{\Gamma_V}{2}(\delta z_{w\bar{w},x+1}+\delta z_{w\bar{w},x-1})+(p-4)\frac{\Gamma_V}{2}\delta z_{w\bar{w},x}.
\end{equation} 
To the leading order of small $V$, an initial perturbation near $x=0$ spreads out as
\begin{equation}
\delta z_{w\bar{w},x}(t)\approx \int \frac{dk}{2\pi}e^{(q-2)\Gamma_Jt-p\Gamma_Vtk^2/2} e^{ikx}\sim e^{(q-2)\Gamma_J t-\frac{x^2}{2\Gamma_V p t}}.
\end{equation}
This give $v_B=\sqrt{2(q-2)p\Gamma_J\Gamma_V}\sim\sqrt{JV}(n(1-n))^{\frac{p+q}{4}-1}$. We compare $v_E^{(m)}$ with $v_B s_0^{(m)}$. At small density $n\approx 0$ or high density $n\approx 1$, we find
\begin{equation}
v_Bs_0^{(m)}\sim (VJ)^{1/2}\frac{mN}{m-1}(n(1-n))^{\frac{p+q}{4}}.
\end{equation}
For small $V$, we thus find $v_B \gg v_E^{(m)}$. The result shows that $v_B$ matches the entanglement velocity $v_E$ only for $p=q$ and $V\sim J$.

Now we consider the entanglement velocity for the Von Neumann entropy $S_A$. In the limit of $m\rightarrow 1$, it is found that additional loop diagrams contribute \cite{dadras2021perturbative}, which cancels the divergence of \eqref{eqn:SmA}. For $m=2$ and $m=4$, the diagrams read 
\begin{equation}
S^{(m)}_A=-\frac{1}{m-1}\log \text{tr}_A(\rho_A^n),\ \ \ \ \ \ S_A=\underset{m\rightarrow 1}{\lim}~S^{(m)}_A=-\text{tr}_A(\rho_A\log \rho_A).
\end{equation}
As an example, for the second and the forth R\'enyi entropy, we have 
\begin{equation}\label{eqn:perturbvon}
\delta S^{(2)}_A\ =\ 
\begin{tikzpicture}[scale=0.65,baseline={([yshift=-5pt]current bounding box.center)}]

\draw[thick,blue] (-70pt,2pt)--(-10pt,2pt);
\draw[thick,blue] (-10pt,-2pt)--(-70pt,-2pt);
\filldraw (-70pt,0pt) circle (2pt) node[left]{\scriptsize 1};
\filldraw (-10pt,0pt) circle (2pt) node[above right]{};

\draw[thick,blue] (10pt,2pt)--(70pt,2pt);
\draw[thick,blue] (70pt,-2pt)--(10pt,-2pt);
\filldraw (10pt,0pt) circle (2pt) node[right]{};
\filldraw (70pt,0pt) circle (2pt) node[right]{\scriptsize 2};

\draw[thick,red] (2pt,-70pt)--(2pt,-10pt);
\draw[thick,red] (-2pt,-10pt)--(-2pt,-70pt);
\filldraw (0pt,-70pt) circle (2pt) node[below]{\scriptsize 2};
\filldraw (0pt,-10pt) circle (2pt) node[above right]{};

\draw[thick,red] (2pt,10pt)--(2pt,70pt);
\draw[thick,red] (-2pt,70pt)--(-2pt,10pt);
\filldraw (0pt,70pt) circle (2pt) node[above]{\scriptsize 1};
\filldraw (0pt,10pt) circle (2pt) node[above right]{};

\draw[thick,dotted] (-50pt,2pt)--(-2pt,50pt);

\draw[thick,dotted] (-30pt,-2pt)--(-2pt,-30pt);

\draw[thick,dotted] (30pt,2pt)--(2pt,30pt);

\draw[thick,dotted] (50pt,-2pt)--(2pt,-50pt);

\filldraw (-50pt,2pt) circle (1.5pt) node[above right]{};
\filldraw (-2pt,50pt) circle (1.5pt) node[above right]{};

\filldraw (-30pt,-2pt) circle (1.5pt) node[above right]{};
\filldraw (-2pt,-30pt) circle (1.5pt) node[above right]{};

\filldraw (30pt,2pt) circle (1.5pt) node[above right]{};
\filldraw (2pt,30pt) circle (1.5pt) node[above right]{};

\filldraw (50pt,-2pt) circle (1.5pt) node[above right]{};
\filldraw (2pt,-50pt) circle (1.5pt) node[above right]{};

\draw[thick] (-30pt,-2pt)--(-50pt,2pt);
\draw[thick] (2pt,30pt)--(-2pt,50pt);
\draw[thick] (30pt,2pt)--(50pt,-2pt);
\draw[thick] (-2pt,-30pt)--(2pt,-50pt);

\filldraw (-40pt,40pt) circle (0pt) node{\scriptsize $V_{ij}^{}$};
\filldraw (40pt,40pt) circle (0pt) node{\scriptsize $V_{kj}^*$};

\filldraw (40pt,-40pt) circle (0pt) node{\scriptsize $V_{kl}^{}$};
\filldraw (-40pt,-40pt) circle (0pt) node{\scriptsize $V_{il}^*$};

\end{tikzpicture}\ \ \ \ \ \ 3\delta S^{(4)}_A\ =\ 
\begin{tikzpicture}[scale=0.65,baseline={([yshift=-5pt]current bounding box.center)}]

\draw[thick,blue] (-70pt,2pt)--(-10pt,2pt);
\draw[thick,blue] (-10pt,-2pt)--(-70pt,-2pt);
\filldraw (-70pt,0pt) circle (2pt) node[left]{\scriptsize 1};
\filldraw (-10pt,0pt) circle (2pt) node[above right]{};

\draw[thick,blue] (10pt,2pt)--(70pt,2pt);
\draw[thick,blue] (70pt,-2pt)--(10pt,-2pt);
\filldraw (10pt,0pt) circle (2pt) node[right]{};
\filldraw (70pt,0pt) circle (2pt) node[right]{\scriptsize 3};

\draw[thick,blue] (2pt,-10pt)--(2pt,-70pt);
\draw[thick,blue] (-2pt,-70pt)--(-2pt,-10pt);
\filldraw (0pt,-70pt) circle (2pt) node[below]{\scriptsize 4};
\filldraw (0pt,-10pt) circle (2pt) node[above right]{};

\draw[thick,blue] (2pt,70pt)--(2pt,10pt);
\draw[thick,blue] (-2pt,10pt)--(-2pt,70pt);
\filldraw (0pt,70pt) circle (2pt) node[above]{\scriptsize 2};
\filldraw (0pt,10pt) circle (2pt) node[above right]{};

\draw[thick,red] (-5.65685pt,8.48528pt)--(-48.0833pt,50.9117pt);
\draw[thick,red] (-50.9117pt,48.0833pt)--(-8.48528pt,5.65685pt);
\filldraw (-49.4975pt,49.4975pt) circle (2pt) node[above left]{\scriptsize 1};
\filldraw (-7.07107pt,7.07107pt) circle (2pt) node[above right]{};

\draw[thick,red] (48.0833pt,50.9117pt)--(5.65685pt,8.48528pt);
\draw[thick,red] (8.48528pt,5.65685pt)--(50.9117pt,48.0833pt);
\filldraw (49.4975pt,49.4975pt) circle (2pt) node[above right]{\scriptsize 2};
\filldraw (7.07107pt,7.07107pt) circle (2pt) node[above right]{};

\draw[thick,red] (-48.0833pt,-50.9117pt)--(-5.65685pt,-8.48528pt);
\draw[thick,red] (-8.48528pt,-5.65685pt)--(-50.9117pt,-48.0833pt);
\filldraw (-49.4975pt,-49.4975pt) circle (2pt) node[below left]{\scriptsize 4};
\filldraw (-7.07107pt,-7.07107pt) circle (2pt) node[above right]{};

\draw[thick,red] (5.65685pt,-8.48528pt)--(48.0833pt,-50.9117pt);
\draw[thick,red] (50.9117pt,-48.0833pt)--(8.48528pt,-5.65685pt);
\filldraw (49.4975pt,-49.4975pt) circle (2pt) node[below right]{\scriptsize 3};
\filldraw (7.07107pt,-7.07107pt) circle (2pt) node[above right]{};

\draw[thick] (-50pt,2pt)--(-30pt,-2pt);
\draw[thick] (-2pt,30pt)--(2pt,50pt);
\draw[thick] (-36.7696pt,33.9411pt)--(-19.799pt,22.6274pt);
\draw[thick] (33.9411pt,36.7696pt)--(22.6274pt,19.799pt);

\draw[thick] (50pt,-2pt)--(30pt,2pt);
\draw[thick] (2pt,-30pt)--(-2pt,-50pt);
\draw[thick] (36.7696pt,-33.9411pt)--(19.799pt,-22.6274pt);
\draw[thick] (-33.9411pt,-36.7696pt)--(-22.6274pt,-19.799pt);

\draw[thick,dotted] (-50pt,2pt)--(-36.7696pt,33.9411pt);
\filldraw (-50pt,2pt) circle (1.5pt) node[above right]{};
\filldraw (-36.7696pt,33.9411pt) circle (1.5pt) node[above right]{};

\draw[thick,dotted] (-30pt,-2pt)--(-22.6274pt,-19.799pt);
\filldraw (-30pt,-2pt) circle (1.5pt) node[above right]{};
\filldraw (-22.6274pt,-19.799pt) circle (1.5pt) node[above right]{};

\draw[thick,dotted] (30pt,2pt)--(22.6274pt,19.799pt);
\filldraw (30pt,2pt) circle (1.5pt) node[above right]{};
\filldraw (22.6274pt,19.799pt) circle (1.5pt) node[above right]{};

\draw[thick,dotted] (50pt,-2pt)--(36.7696pt,-33.9411pt);
\filldraw (50pt,-2pt) circle (1.5pt) node[above right]{};
\filldraw (36.7696pt,-33.9411pt) circle (1.5pt) node[above right]{};

\draw[thick,dotted] (2pt,-30pt)--(19.799pt,-22.6274pt);
\filldraw (2pt,-30pt) circle (1.5pt) node[above right]{};
\filldraw (19.799pt,-22.6274pt) circle (1.5pt) node[above right]{};

\draw[thick,dotted] (-2pt,-50pt)--(-33.9411pt,-36.7696pt);
\filldraw (-2pt,-50pt) circle (1.5pt) node[above right]{};
\filldraw (-33.9411pt,-36.7696pt) circle (1.5pt) node[above right]{};

\draw[thick,dotted] (2pt,50pt)--(33.9411pt,36.7696pt);
\filldraw (2pt,50pt) circle (1.5pt) node[above right]{};
\filldraw (33.9411pt,36.7696pt) circle (1.5pt) node[above right]{};

\draw[thick,dotted] (-2pt,30pt)--(-19.799pt,22.6274pt);
\filldraw (-2pt,30pt) circle (1.5pt) node[above right]{};
\filldraw (-19.799pt,22.6274pt) circle (1.5pt) node[above right]{};

\filldraw (-60pt,25pt) circle (0pt) node{\scriptsize $V_{ij}^{}$};
\filldraw (-25pt,60pt) circle (0pt) node{\scriptsize $V_{kj}^{*}$};
\filldraw (25pt,60pt) circle (0pt) node{\scriptsize $V_{kl}^{}$};
\filldraw (60pt,25pt) circle (0pt) node{\scriptsize $V_{ml}^{*}$};

\filldraw (60pt,-25pt) circle (0pt) node{\scriptsize $V_{mn}^{}$};
\filldraw (25pt,-60pt) circle (0pt) node{\scriptsize $V_{pn}^{*}$};
\filldraw (-25pt,-60pt) circle (0pt) node{\scriptsize $V_{pq}^{}$};
\filldraw (-60pt,-25pt) circle (0pt) node{\scriptsize $V_{iq}^{*}$};
\end{tikzpicture}
\end{equation}
Here the solid black lines represent the fermion Green's function $G^{s\bar{s}}(t)$. There is also a diagram given by the taking charge conjugation of \eqref{eqn:perturbvon}. As the next step, one needs to take the disorder average over Brownian variables, which leads to contractions between $V_{ij}$. An important observation is that if neglect one of the Green's functions (for example the Green's on the blue contour with $w=1$), the rest part of the diagram, after summation over $m$, can be represented by an auxiliary Schwinger-Dyson equation\footnote{Here we add a factor $(-1)^{m-1}$. This is due to the gauge transform of fermion modes $c_d \rightarrow -c_d$ on $(m-1)$ red contours if we hope to set Green's functions the same as $G^{ss'}$ in \eqref{eqn:BrownianG}. Please see \cite{liu2021non} for the details.}
\begin{equation}
\begin{aligned}
\sum_m(m-1)(-1)^{m-1}\delta S^{(m)}_A&=4N\int_0^t \int_0^{t} dt_1dt_2~G^{du}(t_{12})S_\text{aug}^{ud}(t_{21})
\\&\approx 4Nt \int_{-\infty}^{\infty}\frac{d\omega}{2\pi}G^{du}(\omega)S_\text{aug}^{ud}(\omega),
\end{aligned}
\end{equation}
where diagrammatically we have
\begin{equation}
\begin{aligned}
&S^{s\bar{s}}_{\text{aug}}~=~\begin{tikzpicture}[baseline={([yshift=-1pt]current bounding box.center)}, scale=1.2]
\draw[thick,mid arrow] (-12pt,0pt) -- (-24pt,0pt);
\draw[thick,dashed] (-12pt,0pt)..controls (-8pt,12pt) and (8pt,12pt)..(12pt,0pt);
\draw[thick,mid arrow] (12pt,1pt) -- (-12pt,1pt);
\draw[thick,mid arrow] (12pt,-1pt) -- (-12pt,-1pt);
\draw[thick,mid arrow] (24pt,0pt) -- (12pt,0pt);

\filldraw (-20pt,0pt) circle (0pt) node[above]{\scriptsize $s$};
\filldraw (20pt,0pt) circle (0pt) node[above]{\scriptsize $\bar{s}$};
\filldraw (0pt,0pt) circle (0pt) node[below]{\scriptsize $G_\text{aug}^{s\bar{s}}$};
\end{tikzpicture}~+~
\begin{tikzpicture}[baseline={([yshift=-1pt]current bounding box.center)}, scale=1.2]
\draw[thick,mid arrow] (-12pt,0pt) -- (-24pt,0pt);
\draw[thick,dashed] (-12pt,0pt)..controls (-8pt,12pt) and (8pt,12pt)..(12pt,0pt);
\draw[thick,mid arrow] (12pt,1pt) -- (-12pt,1pt);
\draw[thick,mid arrow] (12pt,-1pt) -- (-12pt,-1pt);
\draw[thick,mid arrow] (36pt,0pt) -- (12pt,0pt);

\filldraw (-20pt,0pt) circle (0pt) node[above]{\scriptsize $s$};
\filldraw (16pt,0pt) circle (0pt) node[above]{\scriptsize $\bar{s}$};
\filldraw (32pt,0pt) circle (0pt) node[above]{\scriptsize $s$};
\filldraw (0pt,0pt) circle (0pt) node[below]{\scriptsize $G_\text{aug}^{s\bar{s}}$};
\filldraw (24pt,0pt) circle (0pt) node[below]{\scriptsize $G^{\bar{s}s}$};

\draw[thick,dashed] (36pt,0pt)..controls (40pt,12pt) and (56pt,12pt)..(60pt,0pt);
\draw[thick,mid arrow] (60pt,1pt) -- (36pt,1pt);
\draw[thick,mid arrow] (60pt,-1pt) -- (36pt,-1pt);
\draw[thick,mid arrow] (72pt,0pt) -- (60pt,0pt);
\filldraw (68pt,0pt) circle (0pt) node[above]{\scriptsize $\bar{s}$};

\filldraw (48pt,0pt) circle (0pt) node[below]{\scriptsize $G_\text{aug}^{s\bar{s}}$};
\end{tikzpicture}~+~...\\
&\begin{tikzpicture}[baseline={([yshift=-1pt]current bounding box.center)}, scale=1.2]
\draw[thick,mid arrow] (12pt,1pt) -- (-12pt,1pt);
\draw[thick,mid arrow] (12pt,-1pt) -- (-12pt,-1pt);
\filldraw (0pt,0pt) circle (0pt) node[below]{\scriptsize $G_\text{aug}^{s\bar{s}}$};
\filldraw (8pt,0pt) circle (0pt) node[above]{\scriptsize $\bar s$};
\filldraw (-8pt,0pt) circle (0pt) node[above]{\scriptsize $s$};
\end{tikzpicture}~=~\begin{tikzpicture}[baseline={([yshift=-1pt]current bounding box.center)}, scale=1.2]
\draw[thick,mid arrow] (-12pt,0pt) -- (-36pt,0pt);

\filldraw (-32pt,0pt) circle (0pt) node[above]{\scriptsize $s$};
\filldraw (-16pt,0pt) circle (0pt) node[above]{\scriptsize $\bar s$};
\filldraw (-24pt,0pt) circle (0pt) node[below]{\scriptsize $G^{s\bar{s}}$};
\end{tikzpicture}~+~\begin{tikzpicture}[baseline={([yshift=-1pt]current bounding box.center)}, scale=1.2]
\draw[thick,mid arrow] (-12pt,0pt) -- (-36pt,0pt);
\draw[thick,dashed] (-12pt,0pt)..controls (-8pt,12pt) and (8pt,12pt)..(12pt,0pt);
\draw[thick,mid arrow] (12pt,1pt) -- (-12pt,1pt);
\draw[thick,mid arrow] (12pt,-1pt) -- (-12pt,-1pt);
\draw[thick,mid arrow] (36pt,0pt) -- (12pt,0pt);

\filldraw (-32pt,0pt) circle (0pt) node[above]{\scriptsize $s$};
\filldraw (-16pt,0pt) circle (0pt) node[above]{\scriptsize $\bar s$};
\filldraw (16pt,0pt) circle (0pt) node[above]{\scriptsize $\bar s$};
\filldraw (32pt,0pt) circle (0pt) node[above]{\scriptsize $\bar{s}$};
\filldraw (0pt,0pt) circle (0pt) node[below]{\scriptsize $G_\text{aug}^{\bar{s}s}$};
\filldraw (-24pt,0pt) circle (0pt) node[below]{\scriptsize $G^{s\bar{s}}$};
\filldraw (24pt,0pt) circle (0pt) node[below]{\scriptsize $G^{s\bar{s}}$};
\end{tikzpicture}~+~...
\end{aligned}
\end{equation}
This leads to 
\begin{equation}
\begin{aligned}
&G_\text{aug}(\omega)^{-1}=\begin{pmatrix}
0&G^{ud}(\omega)\\
G^{du}(\omega)&0
\end{pmatrix}^{-1}-
\begin{pmatrix}
0&\Sigma_\text{aug}^{ud}\\
\Sigma_\text{aug}^{du}&0
\end{pmatrix},\\
&\Sigma_\text{aug}^{s\bar{s}}=\int \frac{d\omega}{2\pi}~\frac{V}{2}G^{s\bar{s}}_\text{aug}(\omega),\ \ \ S_\text{aug}^{s\bar{s}}(\omega)=\Sigma_\text{aug}^{s\bar{s}}+\Sigma_\text{aug}^{s\bar{s}}G^{\bar{s}s}(\omega)S_\text{aug}^{s\bar{s}}(\omega).
\end{aligned}
\end{equation}
In other words, we use $G^{s\bar{s}}$ in \eqref{eqn:BrownianG} as $G_0$, and add the effect of $V$ perturbatively. The solution of $\Sigma_\text{aug}^{s\bar{s}}$ takes the form of 
\begin{equation}
\Sigma_\text{aug}^{ud}=-n\sigma,\ \ \ \ \ \ \Sigma_\text{aug}^{du}=(1-n)\sigma,\ \ \ \ \ \ \text{with:}\ \ \frac{ \Gamma  V}{2} \sqrt{\frac{1}{\Gamma ^2+4   (1-n) n \Gamma\sigma }}=\sigma.
\end{equation} 
This gives 
\begin{equation}\label{eqn:Svonpart}
\begin{aligned}
\sum_m(m-1)&(-1)^{m-1}\delta S^{(m)}_A=-8Nt\frac{n(1-n)}{V}\sigma^2
\\&\approx Nt\left\{-2n(1-n)V+4[n(1-n)]^2\frac{V^2}{\Gamma}-12[n(1-n)]^3\frac{V^3}{\Gamma^2} \right\}+O(V^4).
\end{aligned}
\end{equation}
Since diagrammatically $\delta S^{(m)}_A \propto V^m$, we have $(m-1)\delta S^{(m)}_A=-NC(m)[Vn(1-n)/\Gamma]^m\Gamma$. In particular, $C(1)=2$. Unfortunately, we don't find a way to determine the analytical continuation of $C(m)$ near $m=1$ unambiguously. Combining \eqref{eqn:Svonpart} with \eqref{eqn:SmA}, we find 
\begin{equation}\label{eqn:SA}
S_A(t)=2NVtn(1-n)\log\left(\frac{\Gamma e^{1-C'(1)/2}}{Vn(1-n)}\right).
\end{equation}
Comparing \eqref{eqn:SA} and \eqref{eqn:SmA}, we find an additional factor of $-\log \left[V(1-n)n/\Gamma\right]$. We expect this is general for systems with large local Hilbert space dimensions. The enhancement of $\log{V/\Gamma}$ has also been obtained in \cite{dadras2021perturbative} for couplings with Hermitian operators. We attribute enhancement of $\log n$ at a small density to the different behavior of the maximal entropy:
\begin{equation}
s_0^{}\approx -Nn\log n,\ \ \ \ \ \ s_0^{(m>1)}\approx -\frac{Nmn}{m-1}.
\end{equation}

\subsection{Measurements and entanglement transitions}
In this subsection, we consider adding repeated weak measurement to the cBSYK chain \eqref{eqn:cBSYKH_chain}. The evolution of the system then becomes non-unitary, which can exhibit measurement induced entanglement phase transitions. We derive the effective theory for the transition, with a comparison to its Majorana counterpart \cite{zhang2021emergent,jian2021measurement}. 

 We first introduce measurements into the cBSYK chain. We consider  weak measurements with respect to operator $O$, which is described by Kraus operators:
\begin{equation}
K^0_O=1-{\gamma_O} O^\dagger O+O(\gamma^2),\ \ \ \ \ K^1_O=\sqrt{2\gamma_O}O.
\end{equation}
Here we assume $\gamma \ll 1$. In this section, we focus on the second R\'enyi entropy, and perform forced measurement by post-selection of outcome $0$. For $\gamma_O =\zeta_O \delta t$, the evolution of $\rho\otimes \rho$ due to the measurement then takes the form of imaginary-time evolutions \cite{jian2021measurement}
\begin{equation}
\rho(t+\delta t)\otimes \rho(t+\delta t)\propto e^{-h_I\delta t}\rho(t)e^{-h_I\delta t}\otimes e^{-h_I\delta t}\rho(t)e^{-h_I\delta t},
\end{equation}
with $h_I=\zeta_O O^\dagger O$. Adding contributions from measurements with different $O$ and contributions from the unitary part, the total evolution is governed by the non-Hermitian Hamiltonian \cite{liu2021non}:
\begin{equation}
H_{\text{tot}}(t)=H(t)-iH_I,\ \ \ \ \ \ H_I=\sum_O\zeta_O O^\dagger O.
\end{equation}

For complex fermion model, it is natural to measure with respect to the fermion density operator $n_{i,x}=c^{\dagger}_{i,x}c^{}_{i,x}$. However, choosing $\{O\}=\{n_{i,x}\}$ with $\zeta_O=\zeta$ leads to $H_I=\zeta Q$, which commutes the $H(t)$. As a result, the steady state is always with zero density $n$ for any $\zeta>0$. To construct a model with non-trivial entanglement transition, we instead take $\{O\}=\{n_{i\in \text{odd},x},1-n_{i\in \text{even},x}\}$ and $\zeta_O=\zeta$, as a result, we have 
\begin{equation}
H_I=\zeta \sum_{i,x}(-1)^{x-1}c^\dagger_{i,x}c^{}_{i,x} +\text{cons.}
\end{equation}

With the repeated weak measurements, we consider the same protocol as described in subsection \ref{entanglementsetup}. The R\'enyi entropies can still be computed as in \eqref{eqn:contour}. The only difference is the replacement:
\begin{equation}
\text{tr}\log \left(\eta_s\delta_{ww'}^{ss'}\partial_t-\Sigma^{ss'}_{ww',x}\right)~\rightarrow~\text{tr}\log \left(\delta_{ww'}^{ss'}[\eta_s\partial_t-\zeta(-1)^{x}]-\Sigma^{ss'}_{ww',x}\right).
\end{equation}

As pointed out in \cite{zhang2021emergent,jian2021measurement}, the transitions for the second R\'enyi entropy can be understood as a topological defect of the replicated Keldysh contour, which contains two disconnected copies of the traditional Keldysh contour \eqref{eqn:singlecontoursketch}. A volume-law/area-law entangled phase corresponds to a non-vanishing/vanishing correlation between the $u$ $d$ branches. Following this idea, we first study the Green's functions for the steady state on the single Keldysh contour with $w=1$. The Schwinger-Dyson equation reads  
\begin{equation}\label{eqn:SDequation_measurement}
\begin{aligned}
&\begin{pmatrix}
 \partial_t\mp\zeta-\Sigma^{uu}_{e/o}&-\Sigma^{ud}_{e/o}\\
 -\Sigma^{du}_{e/o}& -\partial_t\mp\zeta-\Sigma^{dd}_{e/o}
 \end{pmatrix}\circ
  \begin{pmatrix}
 G^{uu}_{e/o}&G^{ud}_{e/o}\\
 G^{du}_{e/o}& G^{dd}_{e/o}
 \end{pmatrix}=I,\\
&\ \ \ \ \ \ \Sigma^{ss'}_{e/o}~=~\begin{tikzpicture}[baseline={([yshift=-0pt]current bounding box.center)}, scale=1.2]
\draw[thick,mid arrow] (-24pt,0pt) -- (-15pt,0pt);
\draw[thick,dashed] (-15pt,0pt)..controls (-8pt,16pt) and (8pt,16pt)..(15pt,0pt);
\draw[thick,mid arrow] (-15pt,0pt)..controls (-8pt,10pt) and (8pt,10pt)..(15pt,0pt);
\draw[thick,mid arrow] (-15pt,0pt)..controls (-8pt,-10pt) and (8pt,-10pt)..(15pt,0pt);
\draw[thick,mid arrow] (15pt,0pt) -- (-15pt,0pt);
\draw[thick,mid arrow] (15pt,0pt) -- (24pt,0pt);
\filldraw (0pt,-8pt) circle (0pt) node[below]{\scriptsize $e/o$};
\end{tikzpicture}~+~
\begin{tikzpicture}[baseline={([yshift=-3pt]current bounding box.center)}, scale=1.2]
\draw[thick,mid arrow] (-24pt,0pt) -- (-12pt,0pt);
\draw[thick,dashed] (-12pt,0pt)..controls (-8pt,12pt) and (8pt,12pt)..(12pt,0pt);
\draw[thick,mid arrow] (-12pt,0pt) -- (12pt,0pt);
\draw[thick,mid arrow] (12pt,0pt) -- (24pt,0pt);
\filldraw (0pt,1pt) circle (0pt) node[below]{\scriptsize $o/e$};
\end{tikzpicture}.
 \end{aligned}
\end{equation}
Here $G^{ss'}_{e/o}$ is the Green's function for even/odd sites. We would focus on the non-interacting limit with $J=0$, and comment on the interaction effect finally. For $\zeta<V$, the solution takes the forms of
\begin{equation}
G_{e/o}(\omega)=
\begin{pmatrix}
-i\omega\mp\frac{\zeta}{2}&\frac{V}{2\phi}\sqrt{1-\frac{\zeta^2}{V^2}}\\
-\frac{V\phi}{2}\sqrt{1-\frac{\zeta^2}{V^2}}&i\omega\mp\frac{\zeta}{2}
\end{pmatrix}^{-1},
\end{equation}
or in the time domain:
\begin{equation}
G_{e/o}(t)=\frac{1}{2}\begin{pmatrix}
{\text{sgn}(t)}\mp\frac{\zeta}{V}&-{\phi}^{-1}\sqrt{1-\frac{\zeta^2}{V^2}}\\
\phi\sqrt{1-\frac{\zeta^2}{V^2}}&-{\text{sgn}(t)}\mp\frac{\zeta}{V}
\end{pmatrix}e^{-\frac{V}{2}|t|}.
\end{equation}
Interestingly, there is a free parameter $\phi$, which is not fixed by \eqref{eqn:SDequation_measurement}. This is an analog of the density $n$ for the unitary evolution case, which need to be determined by the initial density matrix $\rho=e^{-\mu Q}/\mathcal{Z}$. The solution of \eqref{eqn:BrownianG} indicates we have $\phi=e^{\mu/2}$. We have checked that this matches the numerical results obtained by using methods elaborated in \cite{chen2020replica,liu2021non}. This solution contains non-trivial correlation between $u$ and $d$ branches, and consequently describes a critical phase with logarithmic entanglement entropy. The correlation vanishes as $\sqrt{1-\zeta/V}$ near $\zeta=V$, indicating a mean-field transition in an area-law entangled phase. The solution for $\zeta\geq V$ reads
\begin{equation}
G_{e/o}(\omega)=
\begin{pmatrix}
-i\omega\mp(\zeta-\frac{V}{2})&0\\
0&i\omega\mp(\zeta-\frac{V}{2})
\end{pmatrix}^{-1},
\end{equation}
which gives
\begin{equation}\label{eqn:saddle_area}
G_{e/o}(t)=\frac{1}{2}\begin{pmatrix}
{\text{sgn}(t)}\mp1&0\\
0&-{\text{sgn}(t)}\mp1
\end{pmatrix}e^{-\frac{2\zeta-V}{2}|t|}.
\end{equation}
This solution indicates all particles are in even sites while all odd sites are empty. The solution contains no additional parameters, which indicates the steady state for the area-law entangled phase is independent of the initial density matrix $\rho$.

We are interested in the effective theory for the transition. We start from the area law phase, and consider the fluctuation of $G^{ss'}_{ww',x}(t,t)$ on the replicated Keldysh contour with two worlds. The saddle-point solution of $G^{ss'}_{ww',x}(t,t)$ is given by two copies of \eqref{eqn:saddle_area} as $G^{ss',0}_{ww',x}(t,t)=\delta_{ww'}G_{x\in \text{even}/\text{odd}}^{ss'}(0)$. For fluctuations 
\begin{equation}
G^{ss'}_{ww',x}(t,t)=G^{ss',0}_{ww',x}(t,t)+\delta g^{ss'}_{ww',x}(t),
\end{equation}
expanding the $G-\Sigma$ action to the quadratic order gives
\begin{equation}
S=
\phi^2\sum_{ww'}\int \frac{d\omega}{2\pi}\frac{dk}{2\pi}~(\delta g^{ud}_{ww',e},\delta g^{ud}_{ww',o})^*.
\begin{pmatrix}
2\zeta-V-i\omega&-V\cos(k)\\
-V\cos(k)&2\zeta-V+i\omega
\end{pmatrix}
.(\delta g^{ud}_{ww',e},\delta g^{ud}_{ww',o})^T.
\end{equation}
Here we focus on correlations between $u$ and $d$ branches with $-\delta g^{du}_{ww',x}=\phi^2(\delta g^{ud}_{w'w,x})^*$, which is relevant for the entanglement transition. We introduce the symmetric and anti-symmetric components $\delta g^{s\bar{s}}_{ww',s/a}=\frac{1}{\sqrt{2}}(\delta g^{s\bar{s}}_{ww',e}\pm \delta g^{s\bar{s}}_{ww',o})$, and expand for small $\omega$ and $k$. This leads to
\begin{equation}
S=
\phi^2\sum_{ww'}\int \frac{d\omega}{2\pi}\frac{dk}{2\pi}~(\delta g^{ud}_{ww',s},\delta g^{ud}_{ww',a})^*.
\begin{pmatrix}
2\zeta-2V+V{k^2}/{2}&-i\omega\\
-i\omega&2\zeta
\end{pmatrix}
.(\delta g^{ud}_{ww',s},\delta g^{ud}_{ww',a})^T.
\end{equation}
Now integrate out the anti-symmetric component, we find
\begin{equation}\label{eqn:S1}
S=
\phi^2\sum_{ww'}\int \frac{d\omega}{2\pi}\frac{dk}{2\pi}\left(2\zeta-2V+\frac{V{k^2}}{2}+\frac{\omega^2}{2\zeta}\right)|\delta g^{ud}_{ww',s}(\omega,k)|^2.
\end{equation}
For $\zeta-V>0$, excitations $\delta g^{ud}_{ww',s}$ have positive energy, while for $\zeta-V<0$, they tend to condense, which leads to a finite correlation between $u$ and $d$. To make the condensate value finite, we need to introduce the quartic term. The symmetry of fields $\delta g^{ud}_{ww',s}$ becomes explicit if we treat $\delta g^{ud}_{ww',s}$ as a matrix field $\bm{\varphi}$ with indices $ww'$. \eqref{eqn:S1} now becomes
\begin{equation}
S=
\phi^2\int dx dt \left\{\frac{1}{2\zeta}\text{tr}(\partial_t\bm{\varphi}^{\dagger}\partial_t\bm{\varphi})+\frac{V}{2}\text{tr}(\nabla\bm{\varphi}^{\dagger}\nabla\bm{\varphi})+(2\zeta-2V)\text{tr}(\bm{\varphi}^{\dagger}\bm{\varphi})\right\}.
\end{equation}
This action is invariant under $U(2)_L\otimes U(2)_R$, where the matrix field transforms according to $\bm{\varphi}\rightarrow u_L \bm{\varphi} u_R^\dagger$. This can be traced back to the $U(2)\otimes U(2)$ invariance of the replicated non-interacting Hamiltonian, similar to the $O(2)\otimes O(2)$ symmetry in the Majorana case \cite{zhang2021emergent}. As a result, there are two natural quartic terms that are consistent with the symmetry:
\begin{equation}
\begin{aligned}
S_{\text{full}}=\int dx dt \Big\{&\frac{\phi^2}{2\zeta}\text{tr}(\partial_t\bm{\varphi}^{\dagger}\partial_t\bm{\varphi})+\frac{V\phi^2}{2}\text{tr}(\nabla\bm{\varphi}^{\dagger}\nabla\bm{\varphi})+(2\zeta-2V)\phi^2\text{tr}(\bm{\varphi}^{\dagger}\bm{\varphi})\\
&+\lambda_1\phi^4\text{tr}(\bm{\varphi}^{\dagger}\bm{\varphi})^2+\lambda_2\phi^4\text{tr}(\bm{\varphi}^{\dagger}\bm{\varphi}\bm{\varphi}^{\dagger}\bm{\varphi})\Big\}.
\end{aligned}
\end{equation}
We expect $\lambda_2>0$, which determines property of the symmetry breaking phase. To see this, we consider the special case with $\bm{\varphi}=\varphi_1 \bm{I}+\varphi_2 \bm{\sigma}_x$. We have
\begin{equation}
\text{tr}(\bm{\varphi}^{\dagger}\bm{\varphi})^2=4(\varphi_1^2+\varphi_2^2)^2,\ \ \ \ \ \ \text{tr}(\bm{\varphi}^{\dagger}\bm{\varphi}\bm{\varphi}^{\dagger}\bm{\varphi})=2(\varphi_1^4+3\varphi_1^2\varphi_2^2+\varphi_1^4).
\end{equation}
As a result, for $\lambda_2>0$, the repulsion between $\phi_1$ and $\phi_2$ is larger than the repulsion within each species, and the energy favors $\phi_1\neq 0$, $\phi_2=0$ or $\phi_1= 0$, $\phi_2\neq0$, depending on the boundary condition. In particular, for two copies of the traditional Keldysh contour, $u$ and $d$ branches are paired up with each world, and we have $\phi_1\neq 0$ and $\phi_2=0$. The residue symmetry group is then given by $u_L=u_R$, and there are Goldstone modes living on $U(2)_L\otimes U(2)_R/U(2)_+$. This is different from the Majorana case, where Goldstone modes live on $O(2)_L\otimes O(2)_R/O(2)_+\sim O(2)$. However, we need to point out that in the large-$N$ limit, only the classical configuration with the lowest energy contributes to the entanglement entropy. Even for the cBSYK chain, such a configuration only contains rotations between $\phi_1$ and $\phi_2$, which can be parametrized by $u_R=I$ and $u_L=e^{i \theta \sigma_y}$. Consequently, the entanglement entropy shows the same critical behavior as in Majorana SYK models to the leading order in $1/N$ expansion. We expect the signature of the charge conservation show up to the next order of $1/N$.

Finally, we comment on the interaction effects. Perturbatively, the interaction $J$ contributes a term
\begin{equation}
\delta S=\sum_{ww'}\int  dxdt~\lambda_q\phi^\frac{q}{2}|\varphi_{ww'}|^q.
\end{equation}
For $q\geq4$, this term breaks $U(2)_{L/R}$ to $Z_2\times U(1)\times U(1)$. In this case, the entanglement entropy excites no Goldstone mode, which leads to a volume-law entangled phase.

\section{Discussions}
In this work, we studied the density dependence of the late-time information scrambling and entanglement dynamics in the cBSYK models. Using the Wightman function on a perturbed background, we derive the effective theory between fermions and scramblons for a single-site cBSYK model, which gives analytical results for the late-time OTOC and operator size distribution to the leading order in $1/N$. We then compute the entanglement velocity of the cBSYK chain using perturbative calculations and derive the effective theory for measurement induced transitions with matrix field $\bm{\varphi}$. 

There are many interesting generalizations of the current work. Firstly, it is interesting to consider the late-time information scrambling of cBSYK chains \cite{Agarwal:2022teo}. This requires a complete solution of \eqref{eqn:GWequationchain} with a generalization of \eqref{eqn:GWres}. Secondly, to the leading order of $1/N$, the entanglement entropy for the cBSYK chain with repeated measurements shows the same scaling as its Majorana counterpart. Recently, there is a proposal \cite{tam2022topological} for considering charge weighted version of the entanglement entropy. Similar ideas may be useful in measurement induced transitions with charge conservations. Finally, it is interesting to study the information scrambling and entanglement dynamics in models with non-Abelian symmetries.

\section*{Acknowledgment}
We thank Xiao Chen, Yingfei Gu, Shao-Kai Jian, and Chunxiao Liu for helpful discussions on related topics.

\appendix

\bibliography{ComplexBrownian.bbl}

\end{document}